\documentclass[aps,prl,twocolumn,superscriptaddress,floatfix,nofootinbib]{revtex4-1}

\usepackage[dotinlabels]{titletoc}
\usepackage{titlesec}

\usepackage{xcolor}
\usepackage{mathrsfs}
\usepackage{amsmath}
\usepackage{amssymb}
\usepackage{bm}
\usepackage{amsfonts}
\usepackage{subfigure}
\usepackage{feynmp}
\usepackage{extarrows}
\usepackage{slashed}
\usepackage{graphicx}
\usepackage{dcolumn}
\usepackage{verbatim}
\usepackage{here}
\usepackage{multirow} 
\allowdisplaybreaks[4]
\usepackage{indentfirst}
\usepackage{isodateo}
\usepackage[low-sup]{subdepth}
\usepackage{isodateo}
\usepackage[CJKbookmarks=true, bookmarksnumbered=true,bookmarksopen=true,]{hyperref}
\hypersetup{colorlinks,%
              linkcolor=blue,
              citecolor=blue,
              urlcolor=blue}
\graphicspath{{figs/}}
\usepackage{soul}
\makeatletter
\newsavebox\myboxA
\newsavebox\myboxB
\newlength\mylenA
\newcommand*\xoverline[2][0.7]{%
\sbox{\myboxA}{$\m@th#2$}%
\setbox\myboxB\null
\ht\myboxB=\ht\myboxA%
\dp\myboxB=\dp\myboxA%
\wd\myboxB=#1\wd\myboxA
\sbox\myboxB{$\m@th\overline{\copy\myboxB}$}
\setlength\mylenA{\the\wd\myboxA}
\addtolength\mylenA{-\the\wd\myboxB}%
\ifdim\wd\myboxB<\wd\myboxA%
\rlap{\hskip 0.8\mylenA\usebox\myboxB}{\usebox\myboxA}%
\else
\hskip -0.5\mylenA\rlap{\usebox\myboxA}{\hskip 0.5\mylenA\usebox\myboxB}%
\fi}

\definecolor{Green}{RGB}{199,238,206}

\newcommand{\eqrefe}{Eq.\eqref}
\newcommand{\beq}{\begin{equation}}
\newcommand{\eeq}{\end{equation}}
\newcommand{\ba}{\begin{array}}
\newcommand{\ea}{\end{array}}
\newcommand{\beqa}{\begin{eqnarray}}
\newcommand{\eeqa}{\end{eqnarray}}
\newcommand{\beqs}{\begin{subequations}}
\newcommand{\eeqs}{\end{subequations}}

\newcommand{\la}{\langle}
\newcommand{\ra}{\rangle}
\newcommand{\fr}[2]{\mbox{$\frac{\,{#1}\,}{#2}$}}
\renewcommand{\rm}{\mathrm}
\def\Re{\mathfrak{Re}}
\def\Im{\mathfrak{Im}}

\def\leqq{\leqslant}
\def\geqq{\geqslant}
\def\({\left(}
\def\){\right)}
\def\[{\left[\,}
\def\]{\,\right]}
\def\LB{\left\{}
\def\RB{\right\}}
\def\nn{\nonumber}
\def\pd{\partial}
\def\pp{\prime}
\def\to{\rightarrow}
\def\ito{\!\rightarrow\!}

\def\over{\overline}

\def\tr{\text{tr}}

\def\MP{M_{\text{Pl}}^{}}

\def\ba{\bar{a}}

\def\bfA{\mathbf{A}}

\def\CC{\mathcal{C}}
\def\D{\mathcal{D}}

\def\bE{\bar{E}}
\def\EE{\mathcal{E}}

\def\bfF{\bold{F}}

\def\FF{\mathcal{F}}

\def\ii{\text{i}}
\def\iii{\text{i}\hspace*{0.3mm}}

\def\La{\mathcal{L}}
\def\bM{\xoverline{\mathcal{M}}}
\def\M{\mathcal{M}}

\def\NN{\mathcal{N}}

\def\mO{\mathcal{O}}

\def\bs{\bar{s}}

\def\Sb{\mathbb{S}}

\def\TT{\mathcal{T}}

\def\vt{\tilde{v}}
\def\VV{\mathcal{V}}

\def\al{\alpha}

\def\be{\beta}
\def\ga{\gamma}
\def\Ga{\Gamma}
\def\ka{\kappa}
\def\ab{\alpha\beta}

\def\mn{\mu\nu}

\def\mnr{\mu\nu\rho}

\def\da{\delta}

\def\ep{\epsilon}
\def\vep{\varepsilon}

\def\ct{c_\theta}
\def\st{s_\theta}
\def\cht{c_{\theta/2}}
\def\sht{s_{\theta/2}}

\def\ctt{c_{2\theta}}
\def\cttt{c_{3\theta}}
\def\ctf{c_{4\theta}}
\def\ctfv{c_{5\theta}}

\def\stt{s_{2\theta}}
\def\sttt{s_{3\theta}}

\def\sz{s_0^{}}
\def\tz{t_0^{}}
\def\uz{u_0^{}}
\def\bsz{\bar{s}_0}

\def\sp{\mathfrak{s}}
\def\epT{\epsilon_{\text{T}}}
\def\epP{\epsilon_{\text{P}}}
\def\epL{\epsilon_{\text{L}}}
\def\epS{\epsilon_{\text{S}}}
\def\epX{\epsilon_{\text{X}}}

\def\mt{\widetilde{m}}
\def\P{\text{P}}

\def\T{\text{T}}

\def\APP{A_{\text{P}}^{}}
\def\ATT{A_{\text{T}}^{}}
\def\AP{A^a_{\text{P}}}
\def\AL{A^a_{\text{L}}}
\def\AT{A^a_{\text{T}}}
\def\AS{A^a_{\text{S}}}

\def\AX{A^a_{\text{X}}}
\def\Ap{A_{\text{P}}}
\def\At{A_{\text{T}}}
\def\tdA{\tilde{A}}
\def\tdAT{\tilde{A}^a_{\text{T}}}

\def\hP{h_{\text{P}^{}}}
\def\ZZ{\mathbb{Z}}
\def\hs{\hspace*{0.3mm}}
\def\hsx{\hspace*{0.5mm}}
\def\hsm{\hspace*{-0.3mm}}
\def\hsmx{\hspace*{-0.5mm}}
\def\hf{\frac{1}{2}}
\def\dd{\text{d}}
\def\End{\end{document}}
\allowdisplaybreaks
\begin{document}

\title{%
Topological Equivalence Theorem and Double-Copy
\\[0.5mm]
for Chern-Simons Scattering Amplitudes}

\author{{\sc Yan-Feng Hang}}
\email[]{yfhang717@gmail.com} 
\affiliation{%
T.\,D.$\hs$\,Lee Institute $\&$ School of Physics and Astronomy,\\
Key Laboratory for Particle Astrophysics and Cosmology,\\
Shanghai Key Laboratory for Particle Physics and Cosmology,\\
Shanghai Jiao Tong University, Shanghai, China}
\author{~{\sc Hong-Jian He}}
\email[]{hjhe@sjtu.edu.cn}
\affiliation{%
T.\,D.$\hs$\,Lee Institute $\&$ School of Physics and Astronomy,\\
Key Laboratory for Particle Astrophysics and Cosmology,\\
Shanghai Key Laboratory for Particle Physics and Cosmology,\\
Shanghai Jiao Tong University, Shanghai, China}
\affiliation{%
Physics Department $\&$ Institute of Modern Physics,
Tsinghua University, Beijing, China; \\
Center for High Energy Physics, Peking University, Beijing, China}
\author{\,{\sc Cong Shen}$^{1,}$}
\email[]{congshen.phys@gmail.com} 
\affiliation{%
Fields and Strings Laboratory, Institute of Physics,\\
Ecole Polytechnique Federale de Lausanne, Switzerland}

\begin{abstract} 	
\noindent
We study the mechanism of topological mass-generation for 3d Chern-Simons gauge theories and propose a brand-new Topological Equivalence Theorem to connect scattering amplitudes of the
physical gauge boson states to that of the transverse states under high energy expansion.\ We prove a {\it general energy cancellation mechanism} for $N$-point physical gauge boson amplitudes, which predicts large  cancellations of
$\hs E^{4-L} \!\ito E^{(4-L) - N}\!$
at any $L$-loop level ($L\!\geqq\! 0$).\
We extend the double-copy approach to construct massive graviton amplitudes and study their structures.\ We newly uncover a series of
{\it strikingly large energy cancellations $E^{12}\!\ito E^1$}
of the tree-level four-graviton scattering amplitude
under high energy expansion and establish a new correspondence
between the two energy cancellations
in the topologically massive Yang-Mills gauge theory
and the topologically massive gravity theory.\
We further study the scattering amplitudes of Chern-Simons
gauge bosons and gravitons in the nonrelativistic limit.
\\[1.5mm]
Journal-ref: \href{https://spj.science.org/doi/10.34133/research.0072}{{\tt Research 6\,(2023)\,0072}}
\end{abstract}

\maketitle	


\noindent
{{\bf 1.\,Introduction~}}
\\[2.5mm]
The (2+1)-dimensional (3d) Chern-Simons (CS) theories naturally realize gauge-invariant  (diffeomo-rphism-invariant) topological mass terms for gauge bosons and gravitons\,\cite{Deser:1981wh}.\
Understanding the underlying mechanism
of such topological mass-generations and
how it determines the structure the massive gauge boson/graviton
scattering amplitudes {\it is important}
for applying the modern quantum field theories to
particle physics and condensed matter physics\,\cite{Deser:1981wh}\cite{Dunne:1998}\cite{Tong:2016kpv}.


In this work, we study the dynamics of topological mass-generation
for the 3d CS gauge and gravity theories\,\cite{Deser:1981wh}.\
The 3d gauge fields can acquire gauge-invariant
topological mass terms \`{a} la Chern-Simons\,\cite{CS} and without invoking the conventional Higgs mechanism\,\cite{Higgs}.\
Adding the 3d CS term will convert the transverse polarization state
of massless gauge boson into
the physical polarization state of massive gauge boson, which
conserves the physical degree of freedom (DoF) of a gauge boson:
$1=1\hs$.\
For this, we propose a {\it conceptually new}
Topological Equivalence Theorem (TET)
to formulate the topological mass-generation
at $S$-matrix level,
which quantitatively connects the scattering amplitudes of the
physical polarization states of massive gauge bosons
to that of the corresponding transverse gauge bosons.\
This differs essentially from the conventional equivalence theorem
(ET)$\hs$\cite{ET-Rev} of the 4d Standard Model and from the Kaluza-Klein (KK) ET for the compactified 5d gauge theories\,\cite{5DYM2002}\cite{5DYM2002b}\cite{5DYM2004} and for the compactified 5d General Relativity\,\cite{Hang:2021fmp}\cite{Hang:2024uny}.


We newly develop a general 3d power counting method to count the
leading energy dependence of scattering amplitudes in
both the topologically massive Yang-Mills (TMYM) theory
and topologically massive gravity (TMG).\
By using the TET identity and power counting method for TMYM theories,
we uncover nontrivial energy cancellations among individual
diagrams in the tree-level $N$-gauge boson amplitudes,
$E^4 \ito E^{4-\hsm N}\hsm$,$\hs$
for $N\!\!\geqq\! 4\hs$.\
We will demonstrate that the TET provides a
{\it general mechanism} to
guarantee such highly nontrivial energy cancellations
for the 3d massive gauge boson scattering amplitudes
as well as the 3d massive graviton scattering amplitudes
(through the double-copy construction).


With these, we extend the conventional double-copy approach and
construct the massive four-graviton amplitudes of the TMG theory
from the corresponding four-gauge boson amplitudes of the TMYM theory.\
The conventional double-copy method of
Bern-Carrasco-Johansson\,(BCJ)\,\cite{BCJ}\cite{BCJ-Rev} applies to massless gauge/gravity theories and was inspired by
the Kawai-Lewellen-Tye (KLT)\,\cite{KLT}\cite{Tye-2010} relation
which connects the product of open string amplitudes to that of
the closed string at tree level.\
Some recent works attempted to extend the double-copy method to
the 4d massive YM versus Fierz-Pauli-like
massive gravity\,\cite{dRGT}\cite{DC-4dx1}\cite{DC-4dx2},
to KK-inspired effective gauge theory with extra global U(1)\,\cite{DC-5dx},
to the compactified 5d KK gauge/gravity
theories\,\cite{Hang:2021fmp}\cite{Hang:2024uny}\cite{Li:2022rel},
and to the compactified KK bosonic string theory\,\cite{Li:2021yfk}.\
There are also studies on the double-copy of 3d SUSY CS theroies
in the massless limit\,\cite{songhe}\cite{ythuang}.\
The recent double-copy studies include the 3d CS gauge theories
with or without matter fields and the 3d covariant color-kinematics
duality\,\cite{DC-3dx}\cite{DC-3dx2}\cite{Gonzalez:2021bes}\cite{Moynihan:2021rwh}.

\vspace*{1mm}

Our extended double-copy construction
of the 3d massive four-graviton amplitude
from the 3d massive four-gauge-boson amplitude at tree level
demonstrates {\it strikingly large energy cancellations,
$\,E^{12}\hsm\ito E^1\hsm$,}\,
in the high-energy graviton amplitude.\
With these we establish a {\it new correspondence}
between the two types of distinctive energy cancellations
in the massive gauge-boson amplitudes
and the massive graviton amplitudes:\
$E^4 \hsm\ito E^0\hs$ in the TMYM theory and
$\,E^{12}\!\to\!E^1$ in the TMG theory.\
Finally, for possible applications to the condensed matter system,
we further study the scattering amplitudes of Chern-Simons
gauge bosons and gravitons in the nonrelativistic limit.

\vspace*{3.5mm}
\noindent
{{\bf 2.\,Topological Mass-Generation for Chern-Simons\\ Gauge Theories\,}}
\\[2.5mm]
The 3d Abelian and non-Abelian CS gauge theories may be
called the topologically massive QED (TMQED) and
the topologically massive YM (TMYM), respectively.\
Their Lagrangians take the following forms:
\beqs
\label{eq:L-MCS-YMCS}
\begin{align}
\label{eq:L-MCS}
&\hspace*{-2.2mm}
\La_{\rm{TMQED}}^{}
=-\fr{1}{4} F_{\mn}^2 \!+\!
\fr{1}{2}\mt\,\vep^{\mnr} A_{\mu} \pd_{\nu} A_{\rho}\hs,
\\[2mm]
\label{eq:L-YMCS}
&\hspace*{-2.2mm}
\La_{\rm{TMYM}} =  -\fr{1}{2}  \tr\bfF_{\!\mn}^2
\!+\! \mt \vep^{\mnr} \tr [ \bfA_{\mu} \hsm
( \pd_{\nu}  \!\hsm-\! \fr{2\ii g}{3}\hsm  \bfA_{\nu} )  \bfA_{\rho} ]\hs,
\end{align}
\eeqs
where
$(\bfF_{\!\mn},\bfA_\mu) \!=\! (F_{\!\mn}^a,A^a_\mu) T^a$
and $T^a$ denotes the generator of the SU($N$) gauge group.\
The matter fields can be further added to the above Lagrangians
when needed.\ We note that in Eq.\eqref{eq:L-MCS-YMCS}
the gauge bosons acquire a topological mass
$\,m\!=\!|\mt|$\, from the CS term, where the ratio
$\,\sp={\mt}/{m}=\pm 1$\,
corresponds to their spin\,\cite{Deser:1981wh}\cite{Dunne:1998}.\
Under a general gauge transformation, the action of TMQED theory is
invariant up to a trivial surface term.\
While for the TMYM theory,
the change of its action will contribute to a phase factor
$e^{\ii 2\pi wn}$,
where $w\!\in\! \ZZ$ represents the winding number
which follows from the homotopy group $\Pi_{3}[\rm{SU}(N)] \!\cong\! \ZZ$
\cite{Tong:2016kpv} and $n$ is the CS level $n\!=\! 4\pi \mt/g^2 \!\in\! \ZZ$\,. This ensures the phase factor $\hs e^{\ii 2\pi wn}=1\hs$.

\vspace*{1mm}

The on-shell gauge field has the plane wave solution
$\,A^{a}_{\mu} \!=\! c^a \ep_\mu(p)e^{-\ii p \cdot  x}\,$
from the equation of motion (EOM), where
$\,p^\mu\!A_\mu^a\!=0$\,. Thus, the polarization vector
obeys the following EOM:
\begin{equation}
\label{eq:PolEOM}
( m\hs \eta^{\mn}\!-\ii\hs\sp\hs
\vep^{\mu\rho\nu}p_{\rho}^{})\,\ep_{\nu}^{}(p) = 0 \,.
\end{equation}
Since the CS term does not add any new field, the physical
degrees of freedom of each gauge field $A^{a}_{\mu}$ is conserved
before and after setting
$m=0$ limit\,\cite{Jackiw:1991}\cite{Pisarski:1985}, i.e.,
$1=1$\,.\ The conservation of the physical degrees of freedom
of $A^{a}_{\mu}$  can be further understood from
analyzing the (2+1)d little group\,\cite{UIR3}\cite{supp}.

\vspace*{1mm}

A 3d massive gauge boson in the rest frame has momentum
$\,\bar{p}^\mu\! =\!(m,\,0,\,0)\hs$, and its
physical polarization vector 
is solved as
$\hs\epP^\mu(\bar{p}) \!=\!
\fr{1}{\sqrt{2\,}\,}(0,\hs 1,\hs -\ii\hs\sp )\hs$.
The $\hs\epP^\mu(\bar{p})\hs$ can be boosted
to $\hs\epP^\mu({p})\hs$ for a general momentum
$\,p^\mu \!=\!E(1,\,\be\st,\,\be\ct)$ \cite{supp}.\
We find that $\hs\epP^\mu({p})\hs$
can be generally decomposed as:
\beqs
\label{eq:Pol-All}
\begin{align}
\label{eq:Pol-epP}
\epP^\mu
&= \fr{1}{\sqrt{2\,}\,}\!\(\epT^\mu +\epL^\mu\)\hsm,
\\
\label{eq:Pol-epTL}
\epT^\mu&=(0 ,\, \ii\hs\sp\hs\ct ,\, -\ii\hs\sp\hs \st) ,~~~
\epL^\mu = \bE (\be ,\, \st ,\, \ct)\hs,
\end{align}
\eeqs
where $\epT^\mu$ ($\epL^\mu$) denotes the transverse
(longitudinal) polarization vector,
$\be\!=\!(1\!-\!\bE^{-2})^{1/2}$, $\bE \!=\!E/m$,
and $(\st,\ct)$ $=\! (\sin\hsm\theta,\cos\hsm\theta)$.
Hence, using \eqrefe{eq:Pol-All},
we can define the on-shell polarization
states of the gauge field $A^a_\mu$\,:
\beqs
\label{eq:A-All}
\begin{alignat}{3}
\label{eq:Ap}
\AP \,&=\, \epP^\mu A_\mu^a &&\,=\,
\fr{1}{\sqrt{2\,}\,}\!\(\AT + \AL\)\hsm,
\\
\label{eq:Ax}
\AX \,&=\, \epX^\mu A_\mu^a
&&\,=\, \fr{1}{\sqrt{2\,}\,}\!\(\AT -\AL\)\hsm,
\\
\label{eq:AS}
\AS \,&=\,  \epS^\mu A_\mu^a
&&\,= \, (p^\mu\hsm /m) A^a_\mu \,,
\end{alignat}
\eeqs
where
$\hs\epS^\mu\hsm =\hsm p^{\mu}/m
 \hsm =\hsm\epL^\mu\hsm - v^\mu\hsmx$
with the residual term
$v^\mu \!=\! \mO(m/E)$, and
$\epP^{}\hsm\cdot\epX^{*}\hsm =\epP^{}\hsm\cdot\hsm\epS^{*}
\!=$ $\epX^{}\cdot\epS^{*} \hsm =\hsm 0$\,.\
We note that the 3d massive gauge boson $A^a_\mu$ has 3 possible states
in total, including 1 physical polarization state $\AP$ and 2 unphysical
polarization states ($\AX,\AS$).\
In contrast, the massless gauge boson
contains 1 physical transverse polarization $\AT$\,
and 2 unphysical polarizations $(\AL,\AS)$
with $\hs\epL^\mu\hsm +\epS^\mu \hsm\propto\hsm p^\mu$.\
We observe that adding the CS term for $A^a_\mu$ field
dynamically generates a new massive physical state
$\AP$ and converts its orthogonal combination
$\AX$ into unphysical state,
whereas the scalar-polarization state $\AS$
remains unphysical because it appears in the function
$\FF^a\!\propto\! \AS\hs$
of the gauge-fixing term:
\begin{equation}
\label{eq:GF}
\La_{\rm{GF}}=-\fr{1}{2\xi}(\FF^a)^2 \,, \quad
\FF^a = \pd^\mu\! A^a_\mu  \,.
\end{equation}
We stress that one cannot naively take massless limit
$m\ito 0$ for the massive CS theory because it causes
the polarization vector $\epL^\mu\!\propto\! E/m\ito \infty\,$
and thus $\,\epP^\mu\ito \infty\,$, {\it which makes
the physical state $\AP\ito\infty$ and thus ill-defined.}
Hence, the current analysis of the dynamics of the
massive CS theory is highly nontrivial, from which we
will establish a brand-new topological equivalence theorem (TET)
in the next section.

{\linespread{1.5}
\begin{table*}[t]
\centering
\begin{tabular}{c||c|c|c|c}
\hline\hline
{\small Amplitude}
& $\times \bs^2$
& $\times \bs^{3/2}$
& $\times \bs$
& $\times \bs^{1/2}$
\\
\hline\hline
$\TT_{cs}$
& \ $8\st \,  \,\CC_s$  \
&  \ $\iii 32\st \,\CC_s$ \
& \  $0$
& \  $-\iii 128\st\,\CC_s$ \
\\ \hline
 \ $\TT_{ct}$
&  \ $ -(5 \!+\! 4 \ct \!-\! \ctt)\,\CC_t$  \
& \  $ -\iii 8(2\st\!-\!\stt) \,\CC_t$  \
& \ $ 8(5 \!+\! 3\ctt)\, \CC_t $  \
& \  $ \iii 32(2\st\!+\!\stt)\,\CC_t$  \
\\ \hline
$\TT_{cu}$
&  \ $ (5 \!-\! 4 \ct \!-\! \ctt) \,\CC_u$  \
&  \ $ -\iii 8(2\st\!+\!\stt)\,\CC_u$
& \  $ -8(5 \!+\! 3\ctt)\, \CC_u$  \
&   \ $ \iii 32(2\st\!-\!\stt)\,\CC_u$  \
\\
\hline \hline
$\TT_s$
&  \ $ -8\st \,\,\CC_s$  \
&  \ $ -\iii 56 \st \,\CC_s$  \
&  \ $ -128\ct\, \CC_s$  \
&  \ $ -\iii 32 \st\,\CC_s$  \
\\
\hline
$\TT_t$
&   \ $ (5 \!+\! 4 \ct \!-\! \ctt)\,\CC_t$  \
& \ $ -\iii 8(\st\!+\!\stt)\,\CC_t$  \
&  \ $ -8(5\!+\!16\ct\!+\!3\ctt)\,\CC_t$  \
&  \ $ -\iii 32(7\st\!+\!\stt)\,\CC_t$  \
\\
\hline
$\TT_u$
&   \ $ -(5 \!-\! 4 \ct \!-\! \ctt) \,\CC_u$  \
&  \ $ -\iii 8(\st\!-\!\stt)\,\CC_u$  \
& \  $ 8(5\!-\!16\ct\!+\!3\ctt)\,\CC_u$  \
& \   $ -\iii 32(7\st\!-\!\stt)\,\CC_u $ \
\\
\hline\hline
Sum
& 0
& 0
& 0
& 0
\\
\hline\hline
\end{tabular}
\caption{\baselineskip 12pt
{Energy cancellations for amplitude
$\TT[4\Ap^a]
=\TT_c^{}\!+\!\TT_s \! +\!\TT_t \!+\!\TT_u$\,
in the 3d TMYM theory,
where the contribution of the contact channel is
decomposed into three sub-amplitudes according to the
color factors, $\TT_c=\TT_{cs}+\TT_{ct}+\TT_{cu}$.
The energy factors are
$\,{\bs\hsm =\hsm s /m^2}\!=\hsm 4\bE^2\hs$
and $\hs\bE\!=\! E/m \,$, whereas for the angular dependence the notations
are $(s_{n\theta}^{},\hs c_{n\theta}^{})
\!=\!(\sin\hsm n\theta,\hs \cos\hsm n\theta)$.\
A common overall factor $\,({g^2}\!/{128})\,$ in each amplitude is
not displayed for simplicity.}}
\label{tab:1}
\end{table*}
}

\vspace*{3.5mm}
\noindent
{{\bf 3.\,Formulation of Topological Equivalence Theorem}}
\\[2.5mm]
The CS action from Eq.\eqref{eq:L-MCS-YMCS} is gauge-invariant, and using the method of Refs.\,\cite{ET94}\cite{ET-Rev} we can derive a
Slavnov-Taylor-type identity:
\begin{equation}
\label{eq:F-ID}
\la 0| \hs
\FF^{a_1}(p_1^{})
\cdots
\FF^{a_{\hsm N}}(p_{\hsm N}^{})\hs\Phi
\hs |0 \ra \,=\, 0 \,,
\vspace*{-1mm}
\end{equation}
where $\,\FF^a(p_j)\!=\!-\ii p^\mu_j \! A^a_\mu$\,
and the symbol $\Phi$ denotes
any other on-shell physical fields after the
Lehmann-Symanzik-Zimmermann (LSZ) amputation.\
Since the function $\FF^a$ contains only
one single gauge field $A_\mu^a$,
it is straightforward to amputate each external
$\hs\FF^a$ line by the LSZ reduction, where we impose the
on-shell condition $\,p_j^2\!=\!-m^2\hs$
for each external line.\
From Eq.\eqref{eq:Pol-epP} and the relation
$\hs\epS^\mu \!=\epL^\mu\hsmx - v^\mu$
with the residual term
$v^\mu \!=\! \mO(m/E)$,
we deduce the following identity:
\beq
\label{eq:epS}
\epS^\mu \!=\!\sqrt{2}\hs\epP^\mu-\!(\epT^\mu\hsm +v^\mu),\,
\eeq
Using Eqs.\eqref{eq:AS} and \eqref{eq:epS},
we can reexpress the gauge-fixing function
$\FF^a(p) \!=\!-\ii\hs mA_S^a\,$
as follows:
\begin{equation}
\label{eq:F-Omega}
\FF^a(p) \!=\! -\ii \sqrt{2}\hs m (\AP \!- \Omega^a)\hs ,
\hspace*{3mm}
\Omega^a \!=\! \fr{1}{\sqrt{2\,}\,}(\AT \!+\! v^a)\hs ,
\end{equation}
where $\,v^a \!=\! v^\mu A^a_\mu$\,.
Making the LSZ reduction on \eqrefe{eq:F-ID} and
combining it with \eqrefe{eq:F-Omega},
we derive the following TET identity
for the scattering amplitudes:
\beqs
\label{eq:TET-ID}
\begin{align}
\label{eq:TET-IDa}
\hspace*{-1mm}
& \TT [A_{\P}^{a_1^{}} ,\!\cdots\!,A_{\P}^{a_{\hsm N}}, \Phi]
\,=\, \TT [\tdA_{\T}^{a_1^{}} ,\!\cdots\!, \tdA_{\T}^{a_{\hsm N}} \!, \Phi] \!\,+\!\,\TT_{v}^{}\,,
\hspace*{3mm}
\\[-0.5mm]
\label{eq:Tv}
\hspace*{-1mm}
& \TT_{v}^{} \,=\, \sum_{j=1}^N
\TT[\vt^{a_1}  ,\!\cdots\!, \vt^{a_{j}},\tdA_{\T}^{a_{j+1}} ,\!\cdots\!, \tdA_{\T}^{a_{\hsm N}} ,\Phi ] \,,
\end{align}
\eeqs
where $\tdAT\hsm =\!\fr{1}{\sqrt{2\,}\,}A_{\T}^{a}\,$
and $\vt^a \!=\!\fr{1}{\sqrt{2\,}\,}v^a$.\
In the above, the residual term
$\hs\TT_v^{}\hs$ is suppressed by the factor
$\,v^\mu\hsm\!=\hsm\mO (m/E)\!\ll\! 1\hs$
under high energy expansion.\
The TET identity \eqref{eq:TET-IDa} states that the $\AP$-amplitude equals
the corresponding $\AT$-amplitude in the high energy limit.\
We also observe that the right-hand side (RHS) of
\eqrefe{eq:TET-IDa} receives no multiplicative modification factor
at loop level, because both $\AP$ and $\AT$
belong to the same gauge field $A^a_\mu$.\
This feature differs from the conventional
ET\,\cite{ET94}\cite{ET96} for the SM Higgs mechanism.

\vspace*{1mm}

Generalizing the previous power-counting method in 4d theories\,\cite{weinbergPC}\cite{ETPC-97} and in 5d theories\,\cite{Hang:2021fmp}, we derive a new power-counting rule for the 3d CS gauge theories.\
For a given amplitude, we count the energy-dependence with the power:
%
\begin{equation}
\label{eq:DE-TMYM}
D_{\hsm E} \,=\,  (\EE_{\!A_\P}^{} \!-\EE_{v}^{}) + (4 - \EE - \over{\VV}_3^{}) -L  \,,
\\[-0.5mm]
\end{equation}
where $(\EE,\hs\EE_{\!A_\P}^{}\hsm ,\hs\EE_{v}^{})$
denote the numbers of
the external lines, the external physical states $\AP$, and
the external states with $v^\mu$ factor, respectively.\
The $\over{\VV}_3^{}$ is the number of cubic vertices
containing no derivative (which arise from the non-Abelian
CS term) and $L$ stands for the number of loops.\
For the scattering amplitudes of pure gauge bosons ($\AP$)
with the number of external $\AP$ states
$\,\EE\!=\!\EE_{\!\APP}^{}\!\!\!=\!N$ and $\EE_v\!\!=\!0\,$,
we can use Eq.\eqref{eq:DE-TMYM} to deduce
its leading individual contributions to be
of $\mO(E^4)$ at tree level.
For the scattering amplitudes of pure $\AT$ gauge bosons with
the number of external $\AT$ states
$\,\EE\!=\!\EE_{\!\ATT}^{}\!\!=\!N$ and
$\,\EE_{\!\APP}^{}\!\!\!=\!\EE_v\!\!=0\,$,\,
its individual leading contributions scale like
$\,\mO(E^{4-N})$\,.
With these, our TET identity \eqref{eq:TET-IDa}
guarantees the energy cancellation
in the $N$-gauge boson ($\AP$) scattering amplitude
on its left-hand side (LHS):
$E^4 \ito E^{4-N}$.\
This is because on the RHS of Eq.\eqref{eq:TET-IDa}
the pure $N$-gauge boson
$\AT$-amplitude scales as
$\,\mO(E^{\,4-N})$
and the residual term $\,\TT_v$
(with $\EE_v\!\geqq\! 1$) scales no more than
$\,\mO(E^{\,3-N})$.
We can readily generalize this result up to $L$-loop level
and deduce the following {\it energy power cancellations}$\hs$:
\begin{equation}
\label{eq:DE-cancel2}
\Delta D_{\hsm E} \,=\, D_{\hsm E}[N\!\AP] - D_{\hsm E}[N\!\AT] \,=\, N \,.
\end{equation}

For the sake of later analysis,
we also give the power counting rule on
the high-energy leading $E$-dependence
of graviton scattering amplitudes in the TMG theory\,\cite{supp}:
%
\begin{equation}
\label{eq:DE-TMG}
D_{\hsm E} \,=\, 2\EE_{\hP}^{} \!+ (2 + \VV_{\!d3}^{} +L) \,,
\\[-2mm]
\end{equation}
where $\VV_{\!d3}^{}$ denotes the number of vertices containing
3 partial derivatives coming from the
gravitational CS term in \eqrefe{eq:S-TMG}
and $\,\EE_{\hP}^{}\!$ denotes the number of external
physical graviton states $\hP\hs$.

\vspace*{3.5mm}
\noindent
{{\bf 4.\,Massive Gauge Boson Amplitudes and Energy Cancellations\,}\,}
\\[2.5mm]
In this section, we compute explicitly
the four gauge boson scattering amplitudes
$\TT[\Ap^a\Ap^b \!\ito\! \!\Ap^c\Ap^d] (\equiv\! \TT[4\AP])$ and
$\TT[\At^a\At^b \!\ito\! \At^c\At^d] (\equiv\! \TT[4\AT])$
in the 3d TMYM theory.\
They receive contributions from the contact diagram
and the pole diagrams via $(s,t,u)$ channels, as shown in
the first row of Fig.\,\ref{fig:1}.\
Using the power counting rule \eqref{eq:DE-TMYM},
we deduce that the high-energy leading contributions of
$\TT[4\AP]$ and $\TT[4\AT]$
scale like $E^4$ and $E^0$, respectively.\
Hence, using the TET identity \eqref{eq:TET-IDa},
we would predict the exact energy
cancellations at $\mO(E^4, E^3, E^2,E^1)$
in the physical gauge-boson amplitude $\TT[4\AP]$,
because it should match to the leading energy dependence of
$\TT[4\AT]$ on the RHS of the TET identity \eqref{eq:TET-IDa}.

\begin{figure*}[t]
\centering
\hspace*{-5mm}
\includegraphics[height=4cm]{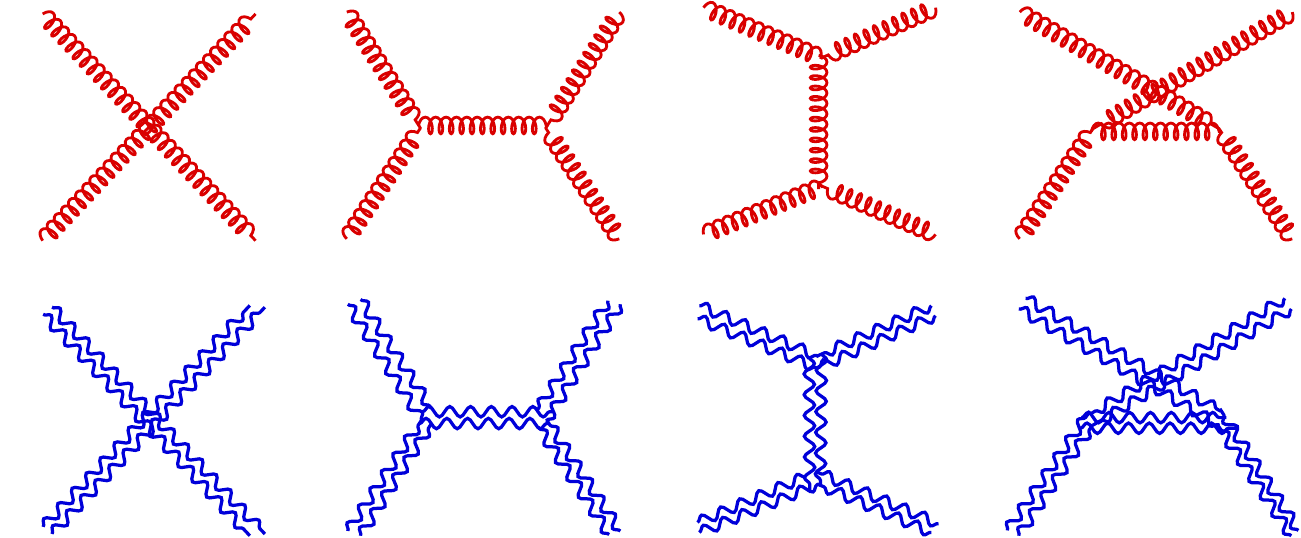}
\caption{\small\baselineskip 11pt
Feynman diagrams in the first row (red color) contribute to
the four-gauge boson scattering processes
$\Ap^a\Ap^b\!\ito\!\Ap^c\Ap^d$\, and
$\,\At^a\At^b\!\ito\!\At^c\At^d$\, in the TMYM theory,
whereas Feynman diagrams in the second row (blue color)
contribute to the four-graviton scattering process
$\hP\hP\ito\hP\hP$ in the TMG theory.
Both the gauge boson and graviton scattering processes
contain the contributions from  	
the contact diagrams and the $(s,t,u)$ channels.}
\label{fig:1}
\vspace*{-5mm}
\end{figure*}

\vspace*{1mm}

Then, we compute the full four-point $\AP$-amplitude at tree level
and present it in the following compact form:
\begin{equation}
\label{eq:4Ap-amp}
\TT[4\AP] \,=\, g^2\! \(\!
\frac{\,\CC_s \,\NN_s\,}{\,s\!-\!m^2\,}
+ \frac{\,\CC_t \,\NN_t\,}{\,t\!-\!m^2\,}
+ \frac{\,\CC_u \,\NN_u\,}{\,u\!-\!m^2\,}  \!\)\!,~~
\end{equation}
where the color factors
$(\CC_s,\hs \CC_t,\hs \CC_u) \hsm =\hsmx
(C^{abe}C^{cde}\!,\hs C^{ade}C^{bce}\!,$ $C^{ace}C^{dbe})$ and
the explicit expressions of kinematic numerators
$(\NN_s,\NN_t,\NN_u)$ are given in
the Supplementary Material\,\cite{supp}.\
We make high energy expansion of the full $\AP$-amplitude
in terms of $\hs 1/\bs\hsx$ or $1/\bsz\hs$,
where $\bs = s/m^2$,
$\bsz\hsm = \sz/m^2$, and $\sz\! =\!s\hsm -\hsm 4m^2$.\
Thus, we can explicitly demonstrate the exact energy cancellations
at each order of $E^n\,(n\!=\!4,3,2,1)$, which are summarized
in Table\,\ref{tab:1}.\
We find that the $\mO(E^4)$ contributions
cancel exactly between the contact diagram and the pole diagram
in each channel of $\hs j\hsm =\hsm s,t,u$\,.\
The sum of each $\mO(E^n)$ contributions (with $n=3,2,1$)
cancels exactly because of the Jacobi identity holds,
$\CC_s\hsm +\CC_t\hsm +\CC_u\!=\!0\hs$.\
For comparison,
we have performed a parallel analysis of the exact energy cancellations
at $\mO(E^n)$ (with $n\!=\!4,3,2,1$) under the high energy expansion
of $1/\bsz$, which are summarized in
the Supplementary Material\,\cite{supp}.\

\vspace*{1mm}

After all the high-energy cancellations,
we systematically derive the leading nonzero
scattering amplitudes of $\TT[4\AP]$ and $\TT[4\AT]$
at $\mO(E^0)$ under the $1/\bs$ expansion,
which take the following forms:
\begin{align}
\label{eq:4AP-4AT-TMYM}
\hspace*{-2mm}
\TT_0[4\AP] \!&= g^2 \!\hsm\[  \!
\CC_s \fr{7\ct}{4} \! +\!
\CC_t \fr{\,7+ 7\ct + 4\ctt}{4(1+\ct)} \!+\!
\CC_u \fr{-(7-7\ct +4\ctt)}{4(1-\ct)} \!\!\]  \!,
\nn\\[1.5mm]
\hspace*{-2mm}
\TT_0[4\AT] \!&= g^2 \!\[  \!
\CC_s \fr{-\ct}{4} \! +\!
\CC_t \fr{3-\ct}{4(1+\ct)} \!+\!
\CC_u \fr{-(3+\ct)\,}{4(1-\ct)} \!\!\]  \!,
\end{align}
where $(\ct,\ctt)\hsm =\hsm (\cos\hsm\theta,\,\cos\hsm 2\theta)$.\
These two amplitudes differ by an amount:
$\TT_0[4\AP]\hsm -\! \hs \TT_0[4\AT]
= 2\ct g^2(\CC_s\hsm +\CC_t \hsm +\CC_u) \!=\! 0$\,,
which vanishes identically due to the Jacobi identity.\
Hence, this demonstrates explicitly that
the TET \eqref{eq:TET-ID} holds in the high energy limit.\
For comparison, we further derive the leading nonzero gauge boson
scattering amplitudes at $\mO(E^0)$
under the $1/\bsz$ expansion:
\begin{align}
\label{eq:4AP-4AT-TMYMs0}
\hspace*{-2mm}
\TT_0'[4\AP] \!&=\hsm g^2 \!\hsm\[  \!
\CC_s \fr{-9\ct}{4} \! +
\CC_t \fr{-(1+ 9\ct + 4\ctt)}{4(1+\ct)} \!+
\CC_u \fr{1-9\ct +4\ctt}{4(1-\ct)} \!\]  \hsm\!,
\nn\\[1.5mm]
\hspace*{-2mm}
\TT_0'[4\AT] \!&= g^2 \!\[  \!
\CC_s \fr{-\ct}{4} \hsm +
\CC_t \fr{3-\ct}{\,4(1+\ct)\,} +
\CC_u \fr{-(3+\ct)\,}{\,4(1-\ct)\,} \!\]  \!.
\end{align}
Inspecting Eqs.\eqref{eq:4AP-4AT-TMYM}-\eqref{eq:4AP-4AT-TMYMs0},
we find that the two amplitudes of transverse gauge bosons
are equal,
$\TT_0[4\AT]\!=\!\TT_0'[4\AT]\hs$,
whereas the two amplitudes of physical
gauge bosons has a difference
$\hs\TT_0'[4\AP]\!-\!\TT_0[4\AP]\hsm =
4\ct g^2(\CC_s\hsm +\CC_t \hsm +\CC_u)\hsm =\hsm 0\hs$,
which vanishes identically due to the Jacobi identity.\
Hence, the leading nonzero amplitudes
of $\mO(E^0)$ are universal and independent of the high-energy
expansion parameters (either $1/\bs$ or $1/\bsz$).\

\vspace*{1mm}

From the above analysis, we have well understood
the structure of the four-gauge boson scattering amplitude
\eqref{eq:4Ap-amp} in the ultraviolet region.\
We have justified its energy cancellations
order by order under the high energy expansion,
at each $\mO(E^n)$ with $n\!=\!4,3,2,1$, and
have proved explicitly
the TET \eqref{eq:TET-IDa} at $\mO(E^0)$.

\vspace*{1mm}

Next, for possible applications to the condensed matter system
and other low energy studies,
we further analyze the nonrelativistic limit and
make the low energy expansion of the four-point gauge boson scattering
amplitudes \eqref{eq:4Ap-amp}.\ Thus, we derive the following
expanded scattering amplitudes of gauge bosons
at the leading order (LO) and next-to-leading order (NLO)
of the low energy expansion:
\\[-5mm]
\beqs
\label{eq:4Ap-NR01}
\begin{align}
\hspace*{-2mm}
\TT_0^{} &= \frac{\,-\iii g^2}{\,4\hs\st\,} e^{\ii 2\theta}
\Big\{\hs\CC_s\hs 2\hs \big[(3\hsm -\hsm 5\ctt)+\iii 5 \stt\big]
\nn\\
\hspace*{-2mm}
&\quad\, + \CC_t \big[(17\!-\hsm 2\hs\ct\hsm -\hsm 19\hs\ctt) +
               \iii (66\hs\st\hsm +\hsm 35\stt)\big]
\\
\hspace*{-2mm}
&\quad\, + \CC_u \big[(17\!+\hsm 2\hs\ct\hsm -\hsm 19\hs\ctt)
- \iii (66\hs\st\!- 35\hs\stt )\big]\!\Big\},
\nn\\
\hspace*{-2mm}
\da\TT &= -\frac{\,\iii g^2\hs s\,}{\,16\hs m^2\st\,}e^{\ii 2\theta}
\Big\{\CC_s\hs 2\hs\big[(-3+5\ctt)\hsm -\iii 5\hs\stt\big]
\nn\\
\hspace*{-2mm}
&\quad\,
+\CC_t \big[\hsm -\hsm (13 +\hsm 2\hs\ct\hsm -\hsm 19\hs\ctt)
 - \iii (46\hs\st\hsm +\hsm 35\hs\stt)\big]
\\
\hspace*{-2mm}
&\quad\,
+\CC_u \big[\hsm -\hsm (13 -\hsm 2\hs\ct\hsm -\hsm 19\hs\ctt) \!
+ \iii (46\hs\st\hsm -\hsm 35 \stt)\big]\Big\} .
\nn
\end{align}
\eeqs
We see that under the nonrelativistic expansion at low energies,
the LO scattering amplitude $\TT_0^{}$ of physical gauge bosons
scales as $E^0$ and the NLO amplitude $\da\TT$ scales as
$E^2\hsm /m^2$.

\vspace*{3.5mm}
\noindent
{{\bf 5.\,Constructing Graviton Scattering Amplitude from Double-Copy}}
\\[2.5mm]
The conventional Einstein gravity in 3d has no physical content\,\cite{Deser:1981wh}\cite{Witten-3dTMG}\cite{Hinterbichler-Rev}, whereas the topologically massive gravity (TMG) includes the gravitational Chern-Simons (CS) action with which the graviton becomes massive and acquires a physical polarization state $\hP\hs$.\
The CS action of the TMG theory  $\int\!\dd^3x\hs\La_{\rm{TMG}}^{}$ contains the Lagrangian:
\begin{equation}
\label{eq:S-TMG}
\hspace*{-2.5mm}
\La_{\rm{TMG}}^{} \!=\! \frac{\,-2~}{\,\ka^2\,}\!\!
\[\!\hsm \sqrt{\hsm -g\hs} R \hsm
-\hsm \frac{\vep^{\mn\hsm\rho}}{2\mt}  \Ga{^\al}_{\hspace{-1.5mm}\rho \be}
\big(\pd_{\mu} \Ga{^\be}_{\hspace{-1.5mm} \al\nu} \!+\!
\frac{2}{3} \Ga{^\be}_{\hspace{-1.5mm}\mu \ga} \Ga{^\ga}_{\hspace{-1.5mm} \nu \al} \big) \!\] \hsmx\!,
\end{equation}
where $\,\ka \!=\!2/\!\sqrt{\MP}\,$
is the gravitational coupling constant and
$\MP$ denotes the Planck mass $\,\MP\!=\!1/(8\pi G)\,$
with $G$ being the Newton constant.

\vspace*{1mm}

We note that the four-point physical gauge boson scattering amplitude
\eqref{eq:4Ap-amp} is invariant under the
generalized gauge transformation:
\begin{equation}
\label{eq:GaugeTrans-N'}
\NN_j^{} \,\to \, \NN_j' = \NN_j^{} +
\Delta\, (s_j^{} \!-\! m^2) \hs ,
\end{equation}
where the index $\hs j\!=\! s,t,u\hs$ and the
coefficient $\Delta$ is an arbitrary function of kinematic variables.\
We find that the numerators $\{\NN_j\}$ of Eq.\eqref{eq:4Ap-amp}
do not manifestly obey
the kinematic Jacobi identity, namely,
$\sum_j \NN_j^{}\!\neq\! 0\,$.\
Then, we require the gauge-transformed numerators
$\{\NN_j'\}$
to satisfy the Jacobi identity $\,\sum_j\hsm\NN_j'\!=\! 0\,$,
with which we determine the coefficient $\Delta\,$
as follows:
\begin{align}
\Delta =&\
-\hsm \frac{\ii}
{\,32\hs m^3\hs\st\,} \hs
\Big[\big(16 m^4s^{-\frac{1}{2}}_{} \!+\!
8 m^2 s^{\frac{1}{2}}_{} \!-\! 3 s^{\frac{3}{2}}_{}\big)
\nn\\
& \
-\hsm\big(16 m^4s^{-\frac{1}{2}}_{}  \!-\!24 m^2 s^{\frac{1}{2}}_{}
-\! 3 s^{\frac{3}{2}}_{}\big)
\ctt \!+\! \iii 16s\hs\st \Big] .
\end{align}
With this, we present the full expressions of the gauge-transformed
numerators $\{\NN_j'\}$ in Eq.(S23) of
the Supplementary Material\,\cite{supp}.\
Thus, from Eq.\eqref{eq:4Ap-amp} we can derive
the following gauge-transformed new scattering amplitude:
\begin{equation}
\label{eq:4Ap-amp-N'}
\TT'[4\AP] \,=\, g^2\! \(\!
\frac{\,\CC_s \,\NN_s'\,}{\,s\!-\!m^2\,}
+ \frac{\,\CC_t \,\NN_t'\,}{\,t\!-\!m^2\,}
+ \frac{\,\CC_u \,\NN_u'\,}{\,u\!-\!m^2\,}  \!\)\!.~~
\end{equation}
We find that each $\NN_j'$ scales as $E^3$ under high energy expansion,
and thus each term in the gauge boson amplitude \eqref{eq:4Ap-amp}
(with numerators given by $\{\NN_j'\}$) should scale as $E^1$.\
Using the gauge-transformed numerators $\{\NN_j'\}$,
the individual terms of the amplitude
\eqref{eq:4Ap-amp-N'} has leading contributions scale as $E^1$
instead of $E^3$.\
We can verify the exact cancellation of the leading $\mO(E^1)$
contributions by summing up them into the following form:
\begin{equation}
\label{eq:4Ap-amp-Nj'-E1}
\hspace*{-2mm}
\TT'[4\AP]_{E^1}^{} =
-\frac{\,\ii\hs g^2\hsm\sqrt{s\,}~}{8\hs m}\!
\frac{\,(7\hsm +\hsm\ctt)\,}{\st}
\big(\CC_s\hsm +\CC_t\hsm +\CC_u\big) =0\,,
\end{equation}
which is proportional to the Jacobi identity and
vanishes identically.\ This cancellation happens in a
similar fashion as the last column of Table\,\ref{tab:1},
but the sum of all terms of last column of Table\,\ref{tab:1}
gives a rather different coefficient (containing distinctive
angular dependence):
\begin{equation}
\label{eq:4Ap-amp-Nj-E1}
\hspace*{-4mm}
\TT [4\AP]_{E^1}^{} =
-\frac{\ii\hs 5\hs g^2\hsm\sqrt{s\,}\,}{4\hs m}\hs{\st}
\big(\CC_s\hsm +\CC_t\hsm +\CC_u\big) =0\,.
\end{equation}
We have further verified that by using the amplitude
\eqref{eq:4Ap-amp-N'} with the gauge-transformed
numerators $\{\NN_j'\}$ and making high energy expansion,
the nonzero leading contribution to the
gauge boson scattering amplitude \eqref{eq:4Ap-amp}
takes the same form as that of Eq.\eqref{eq:4AP-4AT-TMYM}
at $\mO(E^0)$.\ This supports our conclusion that the
leading nonzero gauge boson anmplitudes at $\mO(E^0)$ are
universal, which are independent of the choice of the expansion
parameters (such as $1/\bs$ or $1/\bsz$) and independent of
the basis choice of numerators ($\hs\NN_j^{}$ or $\hs\NN_j'\hs$
as connected by the gauge transformations).

\vspace*{1mm}

Next, we use the power counting rule \eqref{eq:DE-TMG}
to count the leading high-energy dependence of the
four-graviton scattering amplitude at tree level.\
The four-graviton amplitude receives contributions
from the contact diagram and the pole diagrams
via $(s,t,u)$ channels, as shown in
the second row of Fig.\,\ref{fig:1}.\
We find that the leading contributions
of the individual Feynman diagrams to
the physical graviton scattering amplitude scales as $E^{12}$.\
However, using the extended double-copy approach,
we will uncover a series of striking energy-cancellations
in the four-graviton scattering amplitudes,
which make the summation of energy-dependent terms cancel
all the way from $\mO(E^{12})$ down to $\mO(E^1)$.

\vspace*{1mm}

For this purpose, we extend the conventional massless
double-copy method\,\cite{BCJ}\cite{BCJ-Rev}
to the case of TMYM theories.\
Applying the correspondence of the extended color-kinematics duality
$\,\CC_j^{}\ito\NN_j'\,$ to the gauge-transformed four-point
massive gauge boson amplitude \eqref{eq:4Ap-amp-N'},
we construct the scattering amplitude of massive gravitons with
physical polarization,
$\M[\hP^{}\hP\!\ito\hP\hP]\equiv \M[4\hP]$, as follows:
\begin{align}
\label{eq:Amp-4hp-DC}
\M[4\hP] = \frac{\,\ka^2\,}{16}\! \(\!
\frac{{\NN_s^{\,\pp}}^2}{\,s \!-\! m^2\,}
\!+\! \frac{\,{\NN_t^{\,\pp}}^2\,}{\,t \!-\! m^2\,}
\!+\! \frac{\,{\NN_u^{\,\pp}}^2\,}{\,u \!-\! m^2\,}\!\) \!,
\end{align}
where we have made the gauge-gravity coupling
conversion $g \ito\ka/4$\,.\
We stress that, as a key point, the above double-copy construction
{\it must be applied directly to the full gauge-boson amplitude \eqref{eq:4Ap-amp-N'} without high energy expansion.}\
Substituting the numerators
$(\NN_s^{\,\pp},\NN_t^{\,\pp},\NN_u^{\,\pp})$ \cite{supp}
into Eq.\eqref{eq:Amp-4hp-DC}, we derive the following exact
tree-level scattering amplitude of massive gravitons:
\begin{widetext}
\begin{equation}
\label{eq:Amp-4hp-DC3}
\M[4\hP] =
\frac{~\ka^2 m^2
\big(Q_0^{}\hsm +\hsm Q_2^{} \ctt \hsm+\hsm Q_4^{} c_{4\theta}^{}
\hsm +\hsm Q_6^{} c_{6\theta}^{} \hsm +\hsm \bar{Q}_2^{} \stt
\hsm +\hsm \bar{Q}_4^{} s_{4\theta}^{} \hsm +\hsm \bar{Q}_6^{} s_{6\theta}^{}\big)\hsm\csc^2\!\theta~}
{4096\hs\bs^{3/2}\hs (\bs\hsm  -\hsm 1)\hs
[\bs \hsm -\hsm 2 \hsm -\hsm (\bs\hsm -\hsm 4)\ct\hs ]
[\hs \bs \hsm -\hsm 2 \hsm +\hsm (\bs\hsm  -\hsm 4)\ct\hs ]}\,,
\end{equation}
\end{widetext}
where in the numerator the $(Q_j^{},\hs \bar{Q}_j^{})$
are polynomial functions of the dimensionless Mandelstam valiable
$\bs =s/m^2\hs (=\bar{s}_0^{}\!+4)$:
\begin{align}
Q_0^{} &=\hsx
(256 + 49088\hs\bs - 68880\hs\bs^2\!+ 25220\hs\bs^3\! - 2768\hs\bs^4)
\hs\bs^\hf ,
\nn\\
Q_2^{} &=\hsx
(-768 - 45568\hs\bs + 65568\hs\bs^2\!- 19008\hs\bs^3\! +
505\hs\bs^4)\hs\bs^\hf ,
\nn\\
Q_4^{} &=\hsx
4 (192 - 176\hs\bs + 20\hs\bs^2\!+ 635\hs\bs^3\!+ 58\hs\bs^4)\hs\bs^\hf ,
\nn\\
Q_6^{} &=\hsx
-(256 + 2816\hs\bs + 2912\hs\bs^2\!+ 560\hs\bs^3\!+ 17\hs\bs^4)
\hs\bs^\hf,
\nn\\
\bar{Q}_2^{} &=\hsx \ii\hs
(1280 - 256\hs\bs +21312\hs\bs^2\!-8960\hs\bs^3\!+ 475\hs\bs^4)
\hs\bs\hs,
\nn\\
\bar{Q}_4^{} &=\hsx \iii\hs 4
(320 - 544\hs\bs + 676\hs\bs^2\!+ 272\hs\bs^3\!+ 5\hs\bs^4)\hs\bs\hs,
\nn\\
\bar{Q}_6^{} &=\hsx -\ii\hs
(1280 + 3584\hs\bs + 1568\hs\bs^2\! + 128\hs\bs^3\! + \bs^4)
\hs\bs\hs.
\end{align}
The above massive graviton scattering amplitude
can be also reexpressed in terms of
the Mandelstam variable $\sz\, (\hsx =\hsmx s\hsm -\hsm 4m^2)$
which does not contain any mass-dependence,
as shown in Sec.\,4 of the Supplementary Material\,\cite{supp}.

\vspace*{1mm}

Then, we expand Eq.\eqref{eq:Amp-4hp-DC3} by
the high energy expansion of $\hs 1\hsm /s\,$
and derive the four-graviton scattering amplitude at the leading order:
\begin{equation}
\label{eq:Amp0-4hp-DC}
\M_0^{}[4\hP]  =
- \frac{\,\iii \ka^2m\,}{\,2048\,}\hs s^{\hsm\frac{1}{2}}
(494\hs\ct\hsm +\hsm 19\hs\cttt\hsm - c_{5\theta}^{})\hsm
\csc^3 \hspace*{-0.6mm}\theta \hs ,
\end{equation}
which has the distinctive scaling of $\mO(mE)\hs$.\
We can also make the high energy expansion of $\hs 1\hsm /s_0^{}$
{(with $\hs \sz\hsmx =\hsm s\hsm -\hsmx 4m^2$)}
and the LO graviton amplitude takes the same form as
\eqrefe{eq:Amp0-4hp-DC} except the replacement
$s^{1/2}\ito s_0^{1/2}\hs$.

\vspace*{1mm}

From the LO graviton amplitude \eqref{eq:Amp0-4hp-DC},
we can derive its $s$-partial wave
amplitude $a_0^{}$ as follows:
\beqs
\begin{align}
\Re(a_0^{}) &= -\frac{15\ka^2m^2}{\,1024\pi\delta^3\sqrt{s\,}\,}
=\mO\!\(\!\frac{\ka^2m^2}{E}\!\)\!,
\\
\Im(a_0^{}) &=-\frac{247\ka^2m}{\,24576\pi\delta^3\,}
= \mO(\ka^2 mE^0)\hs ,
\end{align}
\eeqs
where we have added an angular cut on the scattering angle
($\delta\hsm\leqq\hsmx\theta\hsm\leqq\!\hsm\pi\!-\!\delta\hs$)
to remove the collinear divergences of the integral.\
We see that the above partial wave amplitudes
have good high energy behaviors
and remain finite in the high energy limit
$E\!\ito\hsm\infty\hs$.\
Imposing the unitarity conditions
$\,|\Re(a_0^{})|\!<\!1/2\,$ and
$\,|\Im(a_0^{})|\hsm\!<\hsm\!1$ \cite{Soldate:1986mk}\cite{He2005},
we deduce the following constraints:
\begin{equation}
\sqrt{s\,} >\frac{\,15\ka^2m^2\,}{\,1024\pi\delta^3\,}\hs,
~~~~m <\frac{\,49152\pi\delta^3\,}{247\ka^2}\,,
\end{equation}
which can be readily obeyed.\
This shows that the 3d TMG exhibits good ultraviolet (UV) behavior,
unlike the conventional 3d Fierz-Pauli-type of massive gravity models.

\vspace*{1mm}

{\linespread{1.75}
\begin{table*}[t]
\centering
\setlength{\tabcolsep}{0.8mm}
\begin{tabular}{c||c|c|c}
\hline\hline
Amplitude
& $\times \bs^2$
& $\times \bs^{3/2}$
& $\times \bs$
\\
\hline\hline
$\M_{s}$
& \ $-\frac{\,99+28 \ctt +\ctf\,}{1-\ctt}$ \
& \  \footnotesize{$-\iii 14(15 \ct  \!+\! \cttt)\! \csc\hsm\theta$} \
&  \ $ \frac{\,2(321 -214 \ctt - 43 \ctf)\,}{1-\ctt} $ \
\\[-5.5mm]
&&&
\\ \hline
$\M_{t}$
& \  $\frac{\,99+28 \ctt +\ctf\,}{4(1-\ct)} $ \
& \  \footnotesize{$\iii (102 \!+\! 105 \ct \!+\! 70 \ctt \!+\!
7 \cttt \!+\! 4 \ctf)\! \csc\hsm\theta$} \
& \  $-\frac{\,321 +559 \ct -214 \ctt -210 \cttt - 43 \ctf - 29 \ctfv}{1-\ctt} $  \
\\[-5.5mm]
&&&
\\ \hline
$\M_{u}$
& \  $\frac{\,99+28 \ctt +\ctf\,}{4(1 + \ct)} $  \
& \   \footnotesize{$\iii (-102 \!+\! 105 \ct \!-\! 70 \ctt
  \!+\! 7 \cttt \!-\! 4 \ctf)\! \csc\hsm\theta$} \
&  \ $-\frac{\,321 - 559 \ct -214 \ctt +210 \cttt - 43 \ctf + 29 \ctfv}{1-\ctt} $ \
\\[-5.5mm]
&&&
\\ 	\hline\hline
Sum
& 0
& 0
& 0
\\
\hline\hline
\end{tabular}
\vspace*{-1mm}
\caption{\baselineskip 12pt
Exact energy cancellations at each order of $(E^4,\,E^3,\,E^2)$
in our double-copied four-graviton scattering amplitude \eqref{eq:Amp-4hp-DC}.	
A common overall factor $\,({\ka^2m^2}\!/{2048})\,$ in each entry
is not displayed for simplicity.}
\label{tab:2}
\end{table*}
}

We note that the leading individual terms of the numerators
$(\NN_j^{},\,\NN_j^{\,\pp})$
scale as $(E^5,\,E^3)$ respectively\,\cite{supp}, 
where the gauge transformation \eqref{eq:GaugeTrans-N'}
causes the energy cancellations of $E^5 \hsm\ito E^3$
in each new numerator $\NN_j^{\,\pp}\hs$.\
This has an important impact on the energy dependence
of the double-copied graviton amplitude \eqref{eq:Amp-4hp-DC}.\
Namely, in each channel, the amplitude
$\,\NN_j^{\,\prime\hs 2}/(s_j^{}\hsm\!-\!m^2)\,$
contains leading energy dependence behaving as $E^4$,
rather than $E^8$ from  $\NN_j^2/(s_j^{}\!-\!m^2)$\,.\
In comparison with the leading energy-dependence of
each individual contribution of
the tree-level four-graviton amplitude
which scales as $E^{12}$ by the direct power counting
of individual Feynman diagrams,
our double-copy construction \eqref{eq:Amp-4hp-DC}
demonstrates that in each channel the graviton scattering amplitude
could have the leading energy dependence of $E^4$ at most.\
Hence, the double-copy construction guarantees
a series of large energy cancellations
in the original four-graviton scattering amplitude,
$E^{12}\ito E^4$,
which brings the leading $E$-dependence down by a large
power factor of $\,E^{8}=E^{4\times 2}$\,.

\vspace*{-1.mm}

In fact, we further discover a series of
striking energy cancellations of
$\,E^{4}\!\ito E^1\hs$
in the full graviton scattering amplitude \eqref{eq:Amp-4hp-DC},
which rely on {\it the sum of all three kinematic channels.}\
We summarize these exact $E$-cancellations in Table\,\ref{tab:2}.
We note that an $S$-matrix element $\,\mathbb{S}\,$ with
$\,\EE\,$ external states and $\hs L\hs$ loops
has mass-dimension
$\,D_\Sb^{} \hsmx =3 -\EE/2\,$
in the 3d spacetime\,\cite{supp}.\
Thus, the four-point graviton scattering amplitude $\M[4\hP]$ in 3d
has mass-dimension 1, and contains
the gravity coupling $\hs\ka^2\hs$
of mass-dimension $-1\hs$.\
Hence, we can express the graviton scattering amplitude
$\,\M[4\hP] \!=\! \ka^2\bM[4\hP]$,\,
where $\bM[4\hP]$
has mass-dimension $2$ and is determined
by the two dimensionful parameters $(E,\,m)$.\
Thus we can deduce its scaling behavior
$\,\bM[4\hP]\!\propto\! m^{n_1} E^{n_2}\,$
with $\,n_1^{}\!\hsm +\hsm n_2^{}\!=\!2\,$,
under the high energy expansion.\
For the energy terms of $\hs E^{n_2^{}}$
with $n_2^{}\!=\!4,3,2\hs$, we deduce its corresponding mass-power factor
$\,n_1^{}\!=\!-2,-1,0\,$, respectively.\
This means that in the massless limit $\,m\hsm\ito 0\,$,
the physical graviton amplitude $\,\bM[4\hP]$\,
would go to infinity (for $n_2^{}\!\hsm\geqq\! 3$) or
remains constant (for $n_2^{}\!=2$\,).
However, we observe that in the massless limit $\,m\hsm\ito 0\,$,
the 3d graviton field becomes unphysical and
has no physical degrees of
freedom\,\cite{Hinterbichler-Rev}.\
Hence the scattering amplitude $\bM[4\hP]$
should vanish since the physical graviton $\hP$ no longer exists
in this limit.\
This means that the $\,m^{n_1} E^{n_2}\,$ terms with
$\,n_1^{}\!=\!-2,-1,0\,$
should vanish and the physical scattering amplitude $\,\bM[4\hP]$
has to start with the leading behavior of
$\hs m^1E^1$, just as the behavior shown in Eq.\eqref{eq:Amp0-4hp-DC}.\
This is why the energy cancellations should hold
at each order of $(E^4,\,E^3,\,E^2)$,
in accord with what we have discovered in Table\,\ref{tab:2}
by the explicit analysis of the energy structure
of the massive graviton amplitude \eqref{eq:Amp-4hp-DC3}.

\vspace*{1mm}

Finally, for possible applications to the condensed matter system
and other low energy studies,
we analyze the nonrelativistic limit and make the low energy
expansion of the double-copied four-graviton scattering
amplitude \eqref{eq:Amp-4hp-DC3}.\ Thus, we derive the following
LO and NLO scattering amplitudes of massive gravitons
under the low energy expansion:
\beqs
\label{eq:4hp-NR01}
\begin{align}
\hspace*{-4mm}
\M_0[4\hP] &= \frac{~\ka^2m^2\,}{\,16\hs s_{\theta}^2\,}
\big(\!-73 + 73 \ctt + \iii 7 \stt\big)\hs e^{\ii 4\theta},
\\
\hspace*{-4mm}
\da\M[4\hP] &= \frac{\,\ka^2s\,}{~64\,s_{\theta}^2~}
\big(55 - 47 \ctt - \iii 17 \stt\big)\hs e^{\ii 4\theta},
\end{align}
\eeqs
where $\,s\hsm\ll\hsm m^2\hs$.\
It shows that under nonrelativistic expansion,
the LO graviton amplitude $\M_0[4\hP]$ scales as
$E^0m^2$ and the NLO graviton amplitude $\da\M[4\hP]$
behaves as $E^2m^0$.

\vspace*{4mm}
\noindent
{{\bf 6.\,Conclusions and Discussions}}
\\[3mm]
Studying the mechanism of topological mass-generations and
its impact on the structure the massive gauge-boson/graviton
scattering amplitudes in the 3d Chern-Simons theories
is important for applying the modern quantum field theories
to particle physics and condensed matter
physics\,\cite{Deser:1981wh}\cite{Dunne:1998}\cite{Tong:2016kpv}.\
In this Letter, we systematically studied the high energy behaviors
of the gauge-boson/graviton scattering amplitudes in the
topologically massive Yang-Mills (TMYM) theory and
the topologically massive gravity (TMG) theory\,\cite{Deser:1981wh}.\
We found that making the high energy expansion uncovers
large energy cancellations
$E^4\ito E^{4-N}$ for each $N$-point
massive gauge boson scattering amplitude.\
These energy cancellations are ensured by the
topological equivalence theorem (TET) identity \eqref{eq:TET-ID}
as we newly proposed in Section\,3.\
This is highly nontrivial
because naively taking the massless limit would cause
the (physical, longitudinal) polarization vectors
in Eq.\eqref{eq:Pol-All} diverge,
$(\epP^{\mu},\hs\epL^{\mu})\ito\infty$,
and thus make the physical state of the topologically massive gauge
boson $\AP$ ill-defined.\
The nontrivial and consistent approach
is {\it to take the high energy expansion for
a fixed nonzero gauge boson mass $m\!\neq\!0$
and prove the large energy cancellations by using the TET identity}
\eqref{eq:TET-ID}, as we demonstrated in Section\,3.\
Moreover, we further extended the conventional massless
double-copy approach to the present massive TMYM and TMG theories.\
We constructed the massive four-graviton
scattering amplitude and uncovered its structure
as in Eqs.\eqref{eq:Amp-4hp-DC}-\eqref{eq:Amp0-4hp-DC}
and Table\,\ref{tab:2}.\
A key point is that
{\it the double-copy construction must be applied
to the exact gauge boson amplitude \eqref{eq:4Ap-amp-N'}
without high energy expansion.}\
From these, we discovered
a series of {\it strikingly large energy cancellations}
in the four-point massive graviton scattering amplitude
at tree level:
\\[-5mm]
\begin{equation}
\label{eq:E12-E1-TMG}
\mO(E^{12}) \,\to\, \mO(E^1)\hs, 
\end{equation}
%
for the 3d TMG theory.\
Our analysis has newly established a {\it striking correspondence between the two types of distinctive energy cancellations} of four-point massive scattering amplitudes:\
$E^4\!\ito E^0$ in the TMYM theory and $E^{12}\!\ito E^1$ in the TMG theory.\
In Eq.\eqref{eq:E12-E1-TMG}, the exact energy cancellations
in the four-graviton scattering amplitude
by a large power of $\,E^{11}$ are even much more severe than the energy cancellations
$E^{10}\!\ito E^2$ in the massive four-longitudinal KK graviton scattering amplitudes of the compactified 5d gravity theory
as found by explicit calculations\,\cite{sekhar}\cite{kurt} and by the KK double-copy construction\,\cite{Hang:2021fmp}\cite{Hang:2024uny}.\
{Our discovery of the striking energy cancellations of $\,E^{12}\ito E^1\hs$ newly demonstrates that the massive graviton scattering amplitudes
in the 3d TMG theory have much better UV behavior
than the naive expectation based on the conventional power counting of Feynman diagrams.\ This also encourages us to further establish
the renormalizability of the TMG theory by extending our massive double-copy approach up to loop levels.\
For the possible applications to the condensed matter system and other low energy studies, we further presented the
nonrelativistic scattering amplitudes of the massive gauge bosons
in Eq.\eqref{eq:4Ap-NR01} and of the massive gravitons in Eq.\eqref{eq:4hp-NR01}.\
A substantial extension of the main content of this Letter is presented in our companion long paper\,\cite{Hang:2021oso}
(where the nonrelativistic scattering amplitudes are not shown).

\newpage
\noindent
{\bf Acknowledgements}\\[1mm]
We thank Stanley Deser and Henry Tye for discussions on this subject.\ This research was supported
in part by the National NSF of China
(under Grants Nos.\,12175136 and 11835005), 
and by National Key R\,\&\,D Program of China (under Grant No.\,2017YFA0402204).


\clearpage
\newpage
\maketitle
\onecolumngrid
\begin{center}
\textbf{\large Topological Equivalence Theorem and Double-Copy
\\[0.5mm]
for Chern-Simons Scattering Amplitudes
\\[3mm]
--- Supplementary Material ---}
\\[5mm]
{\sc Yan-Feng Hang}\,$^1$,\,~ {\sc Hong-Jian He}\,$^{1,2}$,\,~ {\sc Cong Shen}\,$^{1,3}$
\\[2mm]
$^1$\,T.\ D.\ Lee Institute $\&$ School of Physics and Astronomy,\\
Key Laboratory for Particle Astrophysics and Cosmology,\\
Shanghai Key Laboratory for Particle Physics and Cosmology,\\
Shanghai Jiao Tong University, Shanghai, China
\\[1mm] 
$^2$\,Physics Department $\&$ Institute of Modern Physics,
Tsinghua University, Beijing, China;
\\ 
Center for High Energy Physics, Peking University,
Beijing, China
\\[1mm]
$^3$\,Fields and Strings Laboratory, Institute of Physics,
\\
Ecole Polytechnique Federale de Lausanne, Switzerland
\\[1mm]
({yfhang717@gmail.com, hjhe@sjtu.edu.cn, congshen.phys@gmail.com})

\vspace{0.05in}
\end{center}
\setcounter{equation}{0}
\setcounter{figure}{0}
\setcounter{table}{0}
\setcounter{section}{0}
\setcounter{page}{1}
\thispagestyle{empty}
\makeatletter
\renewcommand{\thesection}{S\arabic{section}.}
\renewcommand{\theequation}{S\arabic{equation}}
\renewcommand{\thefigure}{S\arabic{figure}}
\renewcommand{\thetable}{S\arabic{table}}

\vspace*{5mm}

This Supplementary Material provides in detail
the relevant formulas for the analyses of scattering amplitudes
in the 3d Chern-Simons (CS) gauge theory and their double-copy
for the 3d Topologically Massive Gravity (TMG) theory.\
In Section\,S1, we define the kinematics
for the four-point scattering in the 3d spacetime.\
In section\,S2, we derive polarization vectors
for the physical gauge boson states, and present the Feynman rules
of the 3d Topologically Massive Yang-Mills (TMYM) theory.\
In Section\,S3, we present the generalized power counting
method for the scattering amplitudes of the 3d CS gauge theories
and the 3d TMG theory.\
In Section\,S4, we present the complete formulas for
the kinematic numerators of the four-point gauge boson
scattering amplitudes at tree level
(before and after the generalized gauge transformations).\
Then, we present the exact double-copied four-graviton amplitude in terms of the
Mandelstam variable $\sz \,(=\hsm s\hsm -\hsm 4m^2)$ and
its expanded formulas under the high energy expansion
of $\hs m^2\hsmx /\sz\,$.

\section{\hspace*{-10.9cm}S1.~Kinematics of Four-Particle Scattering}
\label{app:A}

The 3d Minkowski
metric tensor and rank-3 Levi-Civita tensor are defined as follows:
\begin{equation}
\eta^{\mn} = \eta_{\mn}=
\begin{pmatrix}
-1 & & \\[-0.5mm]
& 1 & \\[-0.5mm]
&&1
\end{pmatrix}
\hsm\!, \qquad~~~
\vep^{012} = - \vep_{012} = 1 \hs .
\end{equation}
The external momenta for the $2 \ito 2$ elastic scattering process in the center-of-mass frame are given by
\begin{alignat}{3}
\label{eq:CM-Momenta}
p_{1}^\mu \,&=\,  E ( 1, 0,  \be )\hs , \qquad
&&p_2^\mu \,=\,  E ( 1, 0,  -\be )\hs , \quad
\nn\\[-3mm]
\\[-2mm]
p_{3}^\mu \,&=\,   E ( 1, \be \st,  \be \ct )\hs , \qquad
&& p_{4}^\mu \,=\,  E ( 1, -\be \st,  -\be \ct  )\hs ,
\nn
\end{alignat}
where the velocity
$\,\be\!=\!\sqrt{1\!-\!m^2/E^2\,}$\, and
$\,(\st,\ct)\!=\!(\sin \hsm \theta,\, \cos \hsm \theta)\hs$
with $\hs\theta\hs$ being the scattering angle.\
Hence, we can use the momenta \eqref{eq:CM-Momenta} to define
the three Mandelstam variables $\hs (s,t,u)\hs$ as follows:
\begin{align}
s=-\( p_{1} \!+\! p_{2} \)^{2} \!= 4E^2 , ~~~
t=-\( p_{1} \!-\! p_4 \)^{2} \!= -\frac{\,s\,}{2}\be^2
(1\!+\!\ct)\hs ,~~~
u=-\( p_{1} \!-\! p_3 \)^{2} \!=
-\frac{\,s\,}{2}\be^2 (1\!-\!\ct)\hs .
\end{align}
In the present analysis,
we use the on-shell relation
$\,E^2 \!=\! E^2\be^2 + m^2\,$ to
define a new set of mass-independent
Mandelstam variables ($\sz,\tz,\uz$)\,
as follows\,\cite{Hang:2021fmp}:
\begin{align}
\label{eq:s0-t0-u0}
\sz \,=\, 4E^2\be^2   \hs ,  \qquad
\tz \,=\, -\fr{1}{2}\sz(1\!+\!\ct)  \hs , \qquad
\uz \,=\, -\fr{1}{2}\sz(1\!-\!\ct)  \hs ,
\end{align}
where $(s,\,\sz)$ are connected by
$\,\sz \hsm =\hsm s \hsm - 4m^2\hs$.
Furthermore, the summations of $(s,t,u)$ and $(\sz,\tz,\uz)$
satisfy the following relations:
\begin{equation}
s+t+u  = 4 m^{2} , ~~\qquad \sz +\tz + \uz =0 \,.
\end{equation}

\section{\hspace*{-6.6cm}S2.~Polarization States and Feynman Rules
in 3d CS Gauge Theories}
\label{app:B}

The little groups for massless and massive particles in 3d spacetime
are
$\,\ZZ_2^{}\otimes\mathbb{R}$ and
SO(2), respectively\,\cite{UIR3}.\
The 3d Poincar\'e group is ISO(2,1), which
contains the proper Lorentz group
SO(2,1) and the spacetime translations\,\cite{UIR3}\cite{Dunne:1998}.
The 3d Poincar\'e algebra is characterized by two Casimir operators
$(P^2,\,W)=(P_\mu^{}P^\mu,\,P_\mu^{}J^\mu)$,
where $W$ is the Pauli-Lubanski pseudo-scalar and the angular
momentum $J^\mu$ can be generally expressed
as follows\,\cite{Jackiw:1991}:
\begin{equation}
\label{eq:J}
J^\mu \,=\, -\ii\hspace*{0.3mm}\vep^{\mn\al}p_\nu^{}
\frac{\!\partial}{\partial p_\al^{}}
-\sp\frac{~\,p^\mu\!+\!\eta^\mu m~}{\,p\!\cdot\!\eta -m\,}\,,
\end{equation}
with $\,\eta^\mu \!=\!(1,0,0)\,$.\
Thus, the Pauli-Lubanski pseudo-scalar is given by
\,$W\!=P_\mu J^\mu =-\sp\hs m\,$ in the rest frame.\
Hence the spin $\hs\sp\hs$ is also a pseudo-scalar
and takes the values $\,\sp\! =\pm 1$\,
for gauge fields $A^a_\mu$\,.\
The polarization with either
$\,\sp\!=\!+1\,$ or $\,\sp\!=\!-1\,$
is physically equivalent.

\vspace*{1mm}

In the rest frame,
we can solve the equation of motion (2) 
in the main text for the momentum
$\hs\bar{p}^\mu\hsm =(m,\,0,\,0)\hs$:
\begin{equation}
\label{eq:Pol-Rest}
\ep^\mu (\bar{p}) = \fr{1}{\sqrt{2\,}\,}
( 0 ,\, 1 ,\, -\iii \sp )  \,.
\end{equation}
We note that in the rest frame the above gauge boson polarization
vector has zero time-component and its two possible forms
are not independent due to the relation
$\ep_2^\mu = \ii\hspace*{0.25mm}\mathfrak{s}
\hspace*{0.25mm} \ep_1^\mu$.
We can further choose the orthonormal basis
$e^j_1 = (1,0)$ and $e^j_2 = (0,1)$ in a plane,
and define a polarization basis:
\begin{equation}
e^j_\pm \,=\,
\fr{1}{\sqrt{2\,}\,}(e^j_1 \pm \ii\hspace*{0.25mm} e^j_2
\hspace*{0.25mm})
\,=\,
\fr{1}{\sqrt{2\,}\,}\!\(1,\, \pm\ii\hspace*{0.25mm}\) .
\end{equation}
Thus, in the rest frame we can express the spatial components of
the polarization vector $\ep^\mu(\bar{p})$ under the basis
$\{e^j_\pm\}$:
\begin{equation}
\ep^j(\bar{p}) \,=\, \ep_+^{} e^j_+ + \ep_-^{} e^j_- \,,
\end{equation}
where the coefficients $(\ep_+^{},\, \ep_-^{})$ satisfy
$(\ep_+^{},\,\ep_-^{}) = (0,\,1)$ for $\sp=+1$ and
$(\ep_+^{},\,\ep_-^{}) = (1,\,0)$ for $\sp=-1$
\cite{S-Banerjee:2000gc}.
So, as we expect, for the 3d Chern-Simons gauge theory,
the $\,\sp=+1$ case (or, $\sp=-1$ case)
only allows one physical polarization state
$\ep_-^{}$ (or, $\ep_+^{}$)
of the gauge boson.

\vspace*{1mm}

Then, we can make a Lorentz transformation to boost
the polarization vector \eqref{eq:Pol-Rest} in the rest frame
to the following polarization vector
for a general momentum $\,p^\mu \!=\!E(1,\,\be\st,\,\be\ct)\,$:
\begin{equation}
\label{eq:PolP-general}
\epP^\mu(p) \,=\, \frac{1}{\sqrt{2\,}\,}
\!\!\(\!
\frac{\,\iii p_1^{} \hsm\!+\hsm\sp\hs p_2^{}\,}{m} \,, \
\ii\hsm +\!
\frac{\,p_1^{}(\iii p_1^{} \!\hsm +\hsm \sp \hs p_2^{})\,}
{m(m\!-\! p_0^{})} \,,\
\sp\hsm +\!  \frac{\,p_2^{}(\iii p_1^{}\!\hsm +\hsm\sp\hs p_2^{})\,}
{m(m\!-\! p_0^{})}\!\) \!,
\end{equation}
where
$\hs\ep^\mu_{\text{P}+} \!=\! -(\ep^\mu_{\text{P}-})^*\hs$.
Thus, we substitute the momenta \eqref{eq:CM-Momenta} into \eqrefe{eq:PolP-general}, and derive the following explicit form
of the polarization vectors:
\begin{align} 
\ep_{1}^\mu &=\, \frac{\sp}{\sqrt{2}}\! \( \bE\be,\, \iii \sp,\, \bE \)
\!,
\quad~~
\ep_{2}^\mu =\, - \frac{\sp}{\sqrt{2}}
\!\( \bE\be,\, -\iii \sp ,\, -\bE \)\!,
\nn\\
\ep_{3}^\mu &=\, \frac{\sp e^{\ii \sp \theta}}{\sqrt{2}} \!\( \bE\be,\, \bE\st \!+\! \iii \sp \hs \ct,\, \bE\ct \!-\! \iii \sp \hs \st \)
\!,
\quad
\ep_{4}^\mu =\, -\frac{ \sp \hs e^{\ii \sp \theta}}{\sqrt{2}}\!
\( \bE\be,\, -\bE\st \!-\! \iii \sp \hs \ct,\, -\bE\ct
\!+\! \iii \sp \hs \st \) \!,
\label{eq:CSPolarization}
\end{align} 
where we denote a dimensionless energy factor $\bE = E/m$\,.


\vspace*{1mm}

Next, we derive the Feynman rules for the Chern-Simons (CS) gauge theories.
We consider the non-Abelian case of the Topologically Massive Yang-Mills
(TMYM) theory, and the Abelian case corresponds to a special case by
setting all the color indices equal one and
the group structure constant $C^{abc}\!=0\,$.
Thus, the gauge boson propagator for $\sp=\mt/m =+1$ is given by
\begin{align}
\label{eq:prop-TMYM}
\D_{\mn}^{ab}(p) \,=\, -\iii \delta^{ab}\!
\[\! \frac{1}{\,p^2+m^2\,} \!
\(\hsmx \eta_{\mn}^{} - \frac{\,p_{\mu}^{}p_{\nu}^{}\,}{p^2} -
\frac{\,\iii m \vep_{\mnr} p^{\rho}\,}{p^2} \hsm\)
\!+ \xi \frac{\,p_\mu^{}p_\nu^{}\,}{p^4} \!\] \!.
\end{align}
The massless pole of the propagator \eqref{eq:prop-TMYM}
is unphysical because taking $p^2\!=\!0\hs$ with
$m \!\neq\! 0$ in the original
equation of motion\,\cite{Hang:2021oso} leads to
$\,\sp\hs m\hs\vep^{\mu\rho\nu}p_{\rho}^{}\ep_{\nu}^{}\!=\!0\,$
and thus gives the solution of
the polarization vector,
$\ep^\mu \!\propto\! f(p)\hs p^\mu$,
which can be eliminated by the freedom of
gauge transformations\,\cite{Pisarski:1985}.

\vspace*{1mm}

Then, we derive the Feynman rules of cubic and quartic
gauge boson vertices as follows:
\beqs
\begin{align}
V_{\mn\al}^{a b c} \,&= \,
g \hs C^{abc} [ \, \eta_{\mn} (p_{1} \!-\! p_{2})_{\al}
+\eta_{\nu \al}(p_{2}\!-\! p_{3})_{\mu}
+\eta_{\al \mu} (p_{3} \!-\! p_{1})_{\nu}
+ \ii m\vep_{\mn\al} \,] \,,
\\[2mm]
V^{abcd}_{\mn\ab} \,&= \,
- \ii g^{2} \! \[\!\!\!\begin{array}{c}
\quad C^{a b e} C^{c d e}
(\eta_{\mu \al} \eta_{\nu \be}-\eta_{\mu \be} \eta_{\nu \al} )
\\[1mm]
+\, C^{a c e} C^{db e}
( \eta_{\mu \be} \eta_{\nu\al} - \eta_{\mu\nu} \eta_{\al \be} )
\\[1mm]
+\, C^{a d e} C^{b c e}
(\eta_{\mu\nu} \eta_{\al\be}-\eta_{\mu \al} \eta_{\nu\be} )
\end{array} \!\!\]  \!,
\end{align}
\eeqs
where the structure constant appears in the commutator
$[T^a,T^b] = \iii C^{abc} T^c$,
with $T^a$ denoting the generator of the gauge group
SU$(N)\hs$.

\section{\hspace*{-8.2cm}S3.~Power Counting Method for 3d Chern-Simons Theories}
\label{app:C}

Consider a scattering $S$-matrix element $\,\mathbb{S}\,$ having
$\,\EE\,$ external states and $L$ loops ($L\!\geqq\! 0$).
Thus, the amplitude $\,\Sb\,$ has a mass-dimension\,\cite{Hang:2021oso}:
\vspace{-2mm}
\begin{equation}
\label{eq:DS}
D_{\mathbb{S}}^{} \,=\, 3 - \fr{1}{2}\hs\EE  \,,
\end{equation}
where the number of external states
$\,\EE \!=\EE_B^{}+\hs\EE_F^{}\,$
with $\,\EE_B^{}\,$ and $\,\EE_F^{}\,$ being the numbers of
external bosonic and fermionic states, respectively.\
We denote the number of vertices of type-$j$ as $\VV_j^{}$,
where each vertex $\VV_j^{}$ includes $\,d_j^{}\,$ derivatives,
$\,b_j^{}\,$ bosonic lines, and $\,f_j^{}\,$ fermionic lines.\
Then, the total mass-dimension of the energy-independent
coupling constant in the amplitude $\,\mathbb{S}\,$ is given by
\\[-4mm]
\begin{equation}
\label{eq:DC}
D_C^{} \,=\, \sum_j \VV_j^{}\!
\(3-d_j^{}\!- \fr{1}{2} b_j^{}\!- f_j^{}\) \!.
\vspace*{-2mm}
\end{equation}
For each Feynman diagram contributing to the amplitude
\,$\mathbb{S}$\,,\,
we denote the number of the internal lines as
$\,I=I_B^{}+I_F^{}\,$ with
$\,I_B^{}$ ($\,I_F^{}\,$)\, being the number of the internal
bosonic (fermionic) lines. Thus, we have the following general
relations:
\begin{equation}
\label{eq:L-V-I}
L =  1+I-\VV\,, \quad~~
\sum_j \VV_j^{}b_j^{} = 2I_B^{}+\EE_B^{}\,, \quad~~
\sum_j \VV_j^{}f_j^{} = 2I_F^{}+\EE_F^{}\,,
\end{equation}
\\[-3mm]
where $\,\VV\hsm =\hsm\sum_j\!\VV_j^{}\,$
is the total number of vertices in a given Feynman diagram.\
Hence, from Eqs.\eqref{eq:DS}-\eqref{eq:L-V-I},
we derive the leading energy-power dependence of the amplitude
$\,\mathbb{S}\,$ as follows:
%
\begin{equation}
\label{eq:DE0}
D_E^{} \,=\, D_{\mathbb{S}}^{}\! -\!D_C^{}\,=\,
2(1-\VV)+
L+\sum_j  \VV_j^{}\!
\(d_j^{}\!+\!\fr{1}{2}f_j^{}\) .
\end{equation}
\\[-5mm]
Then, we note that the following relations must be obeyed:
\begin{equation}
\sum_j \VV_j^{}d_j^{} =\VV_d^{}\,, \quad
\sum_j \VV_j^{}f_j^{} = 2\VV_F^{}\,, \quad
\VV \!=\!\sum_j\VV_j^{} = \VV_3^{} \!+\! \VV_4^{}\,, \quad
\VV_3^{}\!=\! \VV_d^{} \!+\! \VV_F^{}\!+\! \over{\VV}_3^{}\,,
\vspace*{-1mm}
\end{equation}
where $\hs\VV_d^{}\hs$
denotes the number of all cubic vertices including
one partial derivative
and $\hs\over{\VV}_3^{}\hs$ denotes the number of cubic vertices
without partial derivative\,\cite{Hang:2021fmp}.
With these, we can derive the following power counting rule on
the leading energy dependence:
\begin{equation}
\label{eq:DE}
D_E^{} \,=\, (4 - \EE - \over{\VV}_3^{}) + (\EE_{\hsm \APP}^{} \!-\EE_{v}^{}) -L  \,,
\end{equation}
where $\,\EE_{\hsm \APP}^{}\!$ denotes the number of
external physical gauge bosons $\AP\,(=\epP^\mu A_\mu^a)$
and $\,\EE_{v}^{}\,$ represents the number of
external gauge boson states
contracted with the factor $\hs v^\mu\hs$.

\vspace*{1mm}

For the topologically massive gravity (TMG) theory, we note that
the leading graviton self-interaction vertex comes from the CS term
which always contains 3 partial derivatives. Thus, for a given
graviton vertex of this kind we have
$\hs d_j^{}\hsm = 3\hs$ and $\hs f_j^{}\hsm = 0\hs$
in \eqrefe{eq:DE0}, which lead to
$\,\sum_j\VV_j^{}d_j^{}=3\VV_{d3}^{}\,$ and
$\,\VV =\VV_{d3}^{}\,$ in such leading diagrams,
where $\VV_{d3}^{}$ is the number of vertices including 3 partial
derivatives.  Hence, the leading energy contribution
to the pure graviton scattering amplitude in 3d spacetime
is given by the Feynman diagrams including the CS graviton vertices
with 3 derivatives, and thus is determined as follows:
\\[-5mm]
\begin{equation}
\label{eq:DEhP}
D_E^{} \,=\, 2\hs\EE_{\hP} \!+ (\hs \VV_{d3} +L+2 \hs) \,,
\vspace*{-1mm}
\end{equation}
where $\hs\EE_{\hP}$ denotes the number of external physical graviton states $\hP\,(=\epP^{\mn}h_{\mn})$ and
$\VV_{d3}$ represents the number of vertices including 3
partial derivatives.
For the tree-level diagrams, we have
$\,L\hsm = 0\,$ and $\hs\VV_{d3}^{} = \EE_{\hP} \!\!-\hs 2\,$.
Hence, we can further express the leading energy-power dependence \eqref{eq:DEhP} as follows:
\vspace*{-1.mm}
\begin{equation}
D_E^{0} \,=\, 3 \hs \EE_{\hP} .
\end{equation}

\section{\hspace*{-7.7cm}S4.~Scattering Amplitudes for the TMYM and TMG Theories}
\label{app:D}

For the theories of TMYM and TMG theories, we consider the
scattering processes $\Ap^a\Ap^b\!\ito\!\Ap^c\Ap^d$ $(\At^a\At^b\ito\At^c\At^d)$ and
$\hP \hP \ito \hP\hP$.\
The relevant Feynman diagrams
are presented in Fig.\,1 of the main text.\ 

We systematically derive the kinematic numerators
$(\NN_s,\,\NN_t,\,\NN_u)$
in the four-gauge boson scattering amplitude (13) (main text)
which takes the following form:
\beqs
\label{eq:N-stu}
\begin{align}
\label{eq:N-s}
\NN_s &= \frac{\,4m^2 \!-\! s\,}{\,16 m^3 s^{\frac{1}{2}}_{}\,}
\!\[\! 4 m \hs s^{\frac{1}{2}} (5 m^2 \!+\! 4 s) \ct \!+\! \iii
(4 m^4 \!+\! 29 m^2 s \!+\! 3 s^2 )\st \!\] \!,
\\[1mm]
\label{eq:N-t}
\NN_t &= -\frac{\cht }{\,16m^3\,}\!
\(s^{\frac{1}{2}} \!+\! \iii 2 m
\tan\!\fr{\theta}{2} \)^{\!\!2}\!
\LB 4 m  [  13 m^2\!-\!3 s \!+\! (8m^2 \!-\! s )\ct ]\cht
\!+\! \iii s^{\frac{1}{2}}[22 m^2 \!-\!
3s\!+\!(20 m^2 \!-\! 3 s ) \ct] \sht \RB \!,
\\[1mm]
\NN_u &= \frac{\sht}{\,16m^3\,}\!\(\!s^{\frac{1}{2}}
\!-\!\ii\hs 2m\cot\!\fr{\theta}{2} \!\)^{\!\!2}\!
\LB 4 m  [  13 m^2\!-\!3 s \!-\! (8m^2 \!-\! s )\ct ]\sht
\!-\! \iii s^\hf  [ 22 m^2 \!-\! 3s  \!-\!  (20 m^2 \!-\! 3 s ) \ct]
\cht \RB \!,
\end{align}
\eeqs
where we denote
$(\cht,\,\sht) \!=\! (\sin\hsm\fr{\theta}{2},\,
 \cos\hsm\fr{\theta}{2})\hs$.\
By making the gauge transformation (18) (main text)
on the kinematic numerators $(\NN_s,\,\NN_t,\,\NN_u)\hs$,
we further derive a new set of kinematic numerators
$(\NN_s',\,\NN_t',\,\NN_u')$ which obey the Jabcobi identity
and take following form:
\beqs
\label{eq:N'-stu}
\begin{align}
\NN_s'  =&\ \frac{\ii \hsm\csc \hsm \theta}{8 m \hs s^\hf} \big[ 8 m^4 \!+\! 26 m^2 s
\!-\! 7 s^2 \!-\! (8 m^4 \!+\! 18 m^2 s \!+\! s^2) \ctt \!-\! \iii m \hs s^\hf
(20 m^2 \!+\! 7 s) \stt \big] ,
\\[1mm]
\NN_t' =&\, -\! \frac{\ii \hsm \csc\hsm\theta}{32 m \hs s^\hf}
\big[  (16 m^4 \!+\! 52 m^2 s \!-\! 14 s^2)
\!+\! \(16 m^4 \!+\! 104 m^2 s \!-\! 15 s^2\) \!\ct
\!-\! 2\! \(8 m^4 \!+\! 18 m^2 s \!+\! s^2\) \!\ctt
\nn\\[1mm]
&~+\! \(16 m^4   \!+\! 24 m^2 s   \!+\! s^2\)\! \cttt
+\!\iii ms^\hf(176 m^2 \!+\! 20 s) \st
\!-\! \iii m\hs s^\hf(40 m^2 \!+\! 14 s) \stt
\!-\! \iii m\hs s^\hf(32 m^2 \!+\! 8 s)\sttt \big] ,
\\[1mm]
\NN_u^\pp =&\, -\! \frac{\ii \hsm \csc\hsm\theta}{32m \hs s^\hf}
\big[  (16 m^4 \!+\! 52 m^2 s \!-\! 14 s^2)
\!-\! \(16 m^4 \!+\! 104 m^2 s \!-\! 15 s^2\) \!\ct
\!-\! 2\! \(8 m^4 \!+\! 18 m^2 s \!+\! s^2\) \!\ctt
\nn\\[1mm]
&~+\! \(16 m^4   \!+\! 24 m^2 s   \!+\! s^2\)\! \cttt
-\!\iii m\hs s^\hf(176 m^2 \!-\! 20 s) \st
\!-\! \iii m\hs s^\hf(40 m^2 \!+\! 14 s) \stt
\!+\! \iii m\hs s^\hf(32 m^2 \!+\! 8 s)\sttt \big] ,
\end{align}
\eeqs
where we have defined the notations
$(s_{n\theta}^{},\hs c_{n\theta}^{})
\hsm =\hsm (\sin\hsm n\theta,\, \cos\hsm n\theta)\hs$.\

\vspace*{1mm}

{\linespread{1.5}
\begin{table*}[t]
\centering
\begin{tabular}{c||c|c|c|c}
\hline\hline
{\small Amplitude}
& $\times \bsz^2$
& $\times \bsz^{3/2}$
& $\times \bsz$
& $\times \bsz^{1/2}$
\\
\hline\hline
$\TT_{cs}$
& \ $8\st \,  \,\CC_s$  \
&  \ $\iii 32\st \,\CC_s$ \
& \  $64\ct\,  \CC_s$
& \  $\iii 64\st\,\CC_s$ \
\\ \hline
 \ $\TT_{ct}$
&  \ $ -(5 \!+\! 4 \ct \!-\! \ctt)\,\CC_t$  \
& \  $ -\iii 8(2\st\!-\!\stt) \,\CC_t$  \
& \ $ -32(\ct \!-\!\ctt)\, \CC_t $  \
& \  $ -\iii 16(2\st\!-\!5\stt)\,\CC_t$  \
\\ \hline
$\TT_{cu}$
&  \ $ (5 \!-\! 4 \ct \!-\! \ctt) \,\CC_u$  \
&  \ $ -\iii 8(2\st\!+\!\stt)\,\CC_u$
& \  $ -32(\ct\!+\!\ctt)\,\CC_u$  \
&   \ $ -\iii 16(2\st\!+\!5\stt)\,\CC_u$  \
\\
\hline \hline
$\TT_s$
&  \ $ -8\st \,\,\CC_s$  \
&  \ $ -\iii 56 \st \,\CC_s$  \
&  \ $ -192\ct\, \CC_s$  \
&  \ $ -\iii 368 \st\,\CC_s$  \
\\
\hline
$\TT_t$
&   \ $ (5 \!+\! 4 \ct \!-\! \ctt)\,\CC_t$  \
& \ $ -\iii 8(\st\!+\!\stt)\,\CC_t$  \
&  \ $ -32(3\ct\!+\!\ctt)\,\CC_t$  \
&  \ $ -\iii 16(17\st\!+\!5\stt)\,\CC_t$  \
\\
\hline
$\TT_u$
&   \ $ -(5 \!-\! 4 \ct \!-\! \ctt) \,\CC_u$  \
&  \ $ -\iii 8(\st\!-\!\stt)\,\CC_u$  \
& \  $ -32(3\ct\!-\!\ctt)\,\CC_u$  \
& \   $ -\iii 16(17\st\!-\!5\stt)\,\CC_u $ \
\\
\hline\hline
Sum
& 0
& 0
& 0
& 0
\\
\hline\hline
\end{tabular}
\caption{\baselineskip 12pt
{Energy cancellations for amplitude
$\TT[4\Ap^a]
=\TT_c^{}\!+\!\TT_s \! +\!\TT_t \!+\!\TT_u$\,
in the 3d TMYM theory,
where the contribution of the contact channel is
decomposed into three sub-amplitudes according to the
color factors, $\TT_c=\TT_{cs}+\TT_{ct}+\TT_{cu}$.
The energy factors are
$\,{\bar{s}_0^{}\hsm =\hsm\sz /m^2}\!=\hsm 4\bE^2\!\be^2\hs$
and $\hs\bE\!=\! E/m \,$, whereas for the angular dependence the notations
are $(s_{n\theta}^{},\hs c_{n\theta}^{})
\!=\!(\sin\hsm n\theta,\, \cos\hsm n\theta)$.\
A common overall factor $\,({g^2}\!/{128})\,$ in each amplitude is
not displayed for simplicity.}}
\label{tab:S1}
\vspace*{2mm}
\end{table*}
}

In the main text, we have expanded
the four-point gauge boson scattering amplitudes (13) 
using the above numerators \eqref{eq:N-stu} under the high energy
expansion of $1/\bs$.\ 
We explicitly demonstrated the exact energy cancellations
under the expansion of $1\hsm /\bs\hs$
at each order of $E^n\,(n\!=\!4,3,2,1)$, which are summarized
in Table\,1.\ 
We have also demonstrated the energy cancellations
under the expansion of $1/\bsz$
at each order of $E^n\,(n\!=\!4,3,2,1)$, which we summarize
in the above Table\,\ref{tab:S1}.\
After all these energy cancellations, we have presented the leading
nonzero gauge boson amplitudes in Eq.(15) 
under the under the $1/\bsz$ expansion.\
We find that the amplitude $\TT_0'[4\AP]$ in Eq.(15) 
differs from the amplitude $\TT_0^{}[4\AP]$
in Eq.(14) 
by an amount propotional to
the Jacobi identity, so they are equivalent.\
We also find that the amplitude $\TT_0^{}[4\AT]$
in Eq.(14) 
(under $1\hsm /\bs\hs$ expansion)
and the amplitude $\TT_0'[4\AT]$
in Eq.(15) 
(under $1\hsm /\bsz\hs$ expansion)
are simply equal.\ Hence, the leading nonzero amplitudes
of $\mO(E^0)$ are universal and independent of the high-energy
expansion parameters (either $1/\bs\hs$ or $1/\bsz$).\
Using the gauge-transformed numerators \eqref{eq:N'-stu}
and making high energy expansion,
we have further analyzed the gauge boson scattering
amplitude (20) 
and find that its leading individual terms scale as $E^1$.\
From these, we have demonstrated the remaining energy cancellation
of all the $\mO(E^1)$ terms as shown in Eq.(21)
of the main text, whereas the $\mO(E^1)$ cancellation of the
gauge boson amplitude (13) 
with numerators \eqref{eq:N-stu}
works in a similar fashion but with different
coefficient as shown in Eq.(22) of the main text.

{\linespread{1.75}
\begin{table*}[t]
\centering
\vspace*{2mm}
\begin{tabular}{c||c|c|c}
\hline\hline
Amplitude
& $\times \bsz^2$
& $\times \bsz^{3/2}$
& $\times \bsz$
\\
\hline\hline
$\M_{s}$
& \ $-\frac{\,99+28 \ctt +\ctf\,}{1-\ctt}$ \
& \  \footnotesize{$-\iii 14(15 \ct  \!+\! \cttt)\! \csc\hsm\theta$} \
&  \ $ -\frac{\,2(75 + 326 \ctt +47 \ctf)\,}{1-\ctt} $ \
\\ \hline
$\M_{t}$
& \  $\frac{\,99+28 \ctt +\ctf\,}{4(1-\ct)} $ \
& \  \footnotesize{$\iii (102 \!+\! 105 \ct \!+\! 70 \ctt \!+\!
7 \cttt \!+\! 4 \ctf)\! \csc\hsm\theta$} \
& \  $\frac{\,75 - 107 \ct + 326 \ctt +268 \cttt + 47 \ctf + 31 \ctfv}{1-\ctt} $  \
\\ \hline
$\M_{u}$
& \  $\frac{\,99+28 \ctt +\ctf\,}{4(1 + \ct)} $  \
& \   \footnotesize{$\iii (-102 \!+\! 105 \ct \!-\! 70 \ctt
  \!+\! 7 \cttt \!-\! 4 \ctf)\! \csc\hsm\theta$} \
&  \ $\frac{\,75 + 107 \ct + 326 \ctt -268 \cttt + 47 \ctf - 31 \ctfv\,}{1-\ctt}$ \
\\ 	\hline\hline
Sum
& 0
& 0
& 0
\\
\hline\hline
\end{tabular}
\vspace*{-1mm}
\caption{\baselineskip 12pt
Exact energy cancellations at each order of $(E^4,\,E^3,\,E^2)$
in our double-copied four-graviton scattering amplitude 
(22).	
An common overall factor $\,({\ka^2m^2}\!/{2048})\,$ in each amplitude
is not displayed for simplicity.}
\label{tab:S2}
\vspace*{-5mm}
\end{table*}
}

\vspace*{1mm}

Next, we extend the conventional BCJ double-copy method\,\cite{BCJ}\cite{BCJ-Rev}
to the 3d massive gauge boson and graviton amplitudes,
where we can construct the
desired kinematic numerators \eqref{eq:N'-stu}
in the gauge boson amplitude which obey the Jacobi identity.\
Thus, we apply the color-kinematics duality to the
four-point massive gauge boson scattering amplitude in Eq.(20)
(main text) with the numerators \eqref{eq:N'-stu},
and construct the four-point massive graviton scattering amplitude
of the TMG theory which can be summarized in the following
compact form in terms of the Mandelstam variable
$\sz \,(=s\hsm -\hsm 4m^2)$ and the scattering angle $\theta$\,:
\begin{align}
\M[4\hP]
&\,=\, \frac{\,\ka^2\,}{16}\! \(\!
\frac{{\NN_s'}^2}{\,s \!-\! m^2\,}
\!+\! \frac{\,{\NN_t'}^2\,}{\,t \!-\! m^2\,}
\!+\! \frac{\,{\NN_u'}^2\,}{\,u \!-\! m^2\,}\!\)
\nn\\[1mm]
\label{eq:Amp-4hp-DC2}
&\,=\,
-\frac{~\ka^2 m^2
(P_0^{}\!+\!P_2^{} \ctt \!+\! P_4^{} c_{4\theta}^{} \!+\!
 P_6^{} c_{6\theta}^{} \!+\! \bar{P}_2^{} \stt \!+\! \bar{P}_4^{} s_{4\theta}^{} \!+\! \bar{P}_6^{} s_{6\theta}^{})\hsm\csc^2\!\theta~}
{4096\hs (3 \!+\! \bsz) (4 \!+\! \bsz)^{3/2} (2 \!+\! \bsz \!-\!
 \bsz \ct) (2 \!+\! \bsz \!+\! \bsz \ct)} \,,
\end{align}
where $(P_j,\hs \bar{P}_j)$ are the polynomial functions of
the dimensionless Mandelstam variable $\hsx\bsz =s_0^{}/m^2$,
\begin{align}
\label{eq:Amp-4hp-DC2-Pj}
P_0^{} &=\,-4\hs (7992 \bsz^2 + 4767 \bsz^3 + 692 \bsz^4)
(4\hsm +\hsm \bsz)^\hf \,,
\nn\\[1mm]
P_2^{} &=\, (-221184 - 304128 \bsz - 114048 \bsz^2 - 10928 \bsz^3 + 505 \bsz^4) (4\hsm +\hsm\bsz)^{\hf} ,
\nn\\[1mm]
P_4 &=\, 4\hs (55296 + 45312 \bsz + 13208 \bsz^2 + 1563 \bsz^3 + 58 \bsz^4) (4\hsm +\hsm\bsz)^{\hf} ,
\nn\\[1mm]
P_6 &=\, -(98304 + 57344 \bsz + 11264 \bsz^2 + 832 \bsz^3 + 17 \bsz^4)(4\hsm +\hsm\bsz)^{\hf} ,
\\[1mm]
\bar{P}_2 &=\, \iii (-442368 - 663552 \bsz - 300672 \bsz^2
- 46048 \bsz^3 + 540 \bsz^4 + 475 \bsz^5) \hs,  \hspace*{10mm}
\nn\\[1mm]
\bar{P}_4 &=\, \iii 4\hs (110592 + 104448 \bsz + 36880 \bsz^2 + 5828 \bsz^3 + 372 \bsz^4 + 5 \bsz^5) \hs ,
\nn\\[1mm]
\bar{P}_6 &=\,
-\iii (196608 + 139264 \bsz + 35328 \bsz^2 + 3776 \bsz^3
+ 148 \bsz^4 + \bsz^5) \hs .
\nn
\end{align}
Then, parallel to Eq.(26) 
of the main text,
we make the high energy expansion in terms of $1/\bsz\hs$, and obtain
the following leading nonzero graviton scattering amplitude:
%
\begin{equation}
\label{eq:Amp0-4hp-DC-s0}
\hspace*{-4mm}	
\M_0'[4\hP]  =
- \frac{\,\iii \ka^2m\,}{\,2048\,}\hs s_0^{\hsm\frac{1}{2}}
(494\hs\ct\hsm +\hsm 19\hs\cttt\hsm - c_{5\theta}^{})\hsm
\csc^3 \hspace*{-0.6mm}\theta \hs ,
\end{equation}
which scales as $\mO(mE)\hs$ and takes the same form as
Eq.(26) 
except the replacement
$s^{1/2}\!\ito s_0^{1/2}\hs$.\
We have further verified the exact energy cancellations
at each order of $\bsz^{n/2}$ with $\hs n\!=\!4,3,2$,
which are summarized by the present Table\,\ref{tab:S2}
in parallel to the Table\,2 
of the main text.


\begin{thebibliography}{99}

\bibitem{Deser:1981wh}
S.~Deser, R.~Jackiw, and S.~Templeton,
Phys.\ Rev.\ Lett.\ 48 (1982) 975;
Annals Phys. 140 (1982) 372-411.


\bibitem{Dunne:1998}
For a review, e.g., G.~V.~Dunne,
``Aspects of Chern-Simons theory'',
arXiv:hep-th/9902115 [hep-th].


\bibitem{Tong:2016kpv}
D.~Tong,
``Lectures on the Quantum Hall Effect'',
arXiv:1606.06687 [hep-th].


\bibitem{CS}
E.g., S.\ S.\ Chern, ``Complex Manifolds
without Potential Theory'', second edition,
Springer, Berlin, 1979.


\bibitem{Higgs}
F.\ Englert and R.\ Brout, Phys.\ Rev.\ Lett.\ 13 (1964) 321;
P.\ W.\ Higgs, Phys.\ Rev.\ Lett.\ 13 (1964) 508;
Phys.\ Lett.\ 12 (1964) 132;
G.\ S.\ Guralnik, C. R. Hagen and T. Kibble,
Phys.\ Rev.\ Lett.\ 13 (1965) 585;
T. Kibble, Phys.\ Rev.\ 155 (1967) 1554.




\bibitem{ET-Rev}
For a comprehensive review of the 4-dimensional ET, see:
H.\ J.\ He, Y.\,P.\ Kuang, C.\,P.\ Yuan,
arXiv:hep-ph/9704276 and DESY-97-056, in the proceedings of
CCAST Workshop on ``Physics at the TeV Energy Scale'', vol.72, p.119
(1996).


\bibitem{5DYM2002}
R.\ S.\ Chivukula, D.\,A.\ Dicus, H.\ J.\ He,
Phys.\ Lett.\ B 525 (2002) 175 [hep-ph/0111016].


\bibitem{5DYM2002b}
R.\ S.\ Chivukula and H.\ J.\ He,
Phys.\ Lett.\ B 532 (2002) 121 [arXiv:hep-ph/0201164];
R.\ S.\ Chivukula, D.\ A.\ Dicus, H.\ J.\ He, S.\ Nandi,
Phys.\ Lett.\ B 562 (2003) 109 [hep-ph/0302263].


\bibitem{5DYM2004}
H.-J. He, Int.\ J.\ Mod.\ Phys.\ A\,20 (2005) 3362
[arXiv:hep-ph/0412113],
(in its section\,3), and
presentation at DPF-2004:\,Annual Meeting of the Division of Particles
and Fields, American Physical Society, August 26-31, 2004,
Riverside, California, USA.


\bibitem{Hang:2021fmp}
Y.-F.~Hang and H.-J.~He,
Phys.\ Rev.\ D 105 (2022) 084005 [arXiv:2106.04568 [hep-th]];
Research 2022 (2022) 9860945 [arXiv:2207.11214 [hep-th]].


\bibitem{Hang:2024uny}
Y.-F.~Hang, W.-W. Zhao, H.-J.~He, and Y.-L. Qiu, 
arXiv: 2406.12713 [hep-th].


\bibitem{BCJ}
Z.\ Bern, J.\,J.\ M.\ Carrasco, and H.\ Johansson,
Phys.\ Rev.\ D 78 (2008) 085011
[arXiv:{0805.3993} [hep-th]];
Phys.\ Rev.\ Lett. 105 (2010) 061602
[arXiv:{1004.0476} [hep-th]].


\bibitem{BCJ-Rev}
For a review,
Z.\ Bern, J.\ J.\ M.\ Carrasco, M.\ Chiodaroli, H.\ Johansson, and R.\ Roiban,
arXiv:{1909.01358} [hep-th].


\bibitem{KLT}
H.\ Kawai, D.\,C.\ Lewellen, and S.\,H.\,H.\ Tye,
Nucl.\ Phys.\ B\,269 (1986) 1-23.


\bibitem{Tye-2010}
S.\,H.\,H.\ Tye and Y.\ Zhang,
JHEP 06 (2010) 071 [arXiv: 1003.1732 [hep-th]].


\bibitem{dRGT}
C.\ de\,Rham and G.\ Gabadadze,
Phys.\ Rev.\ D\,82 (2010) 044020 [arXiv:1007.0443 [hep-th]];
C.\ de Rham, G.\ Gabadadze, and A.\ J.\ Tolley,
Phys.\ Rev.\ Lett.\ 106 (2011) 231101
[arXiv:1011.1232 [hep-th]].


\bibitem{DC-4dx1}
A.\ Momeni, J.\ Rumbutis, A.\ J.\ Tolley,
JHEP 12 (2020) 030 [2004.07853 [hep-th]].


\bibitem{DC-4dx2} 
L.\ A.\ Johnson, C.\,R.\,T.\ Jones, and S.\ Paranjape,
JHEP 02 (2021) 148 [arXiv:2004.12948 [hep-th]].

\bibitem{DC-5dx}
A.\ Momeni, J.\ Rumbutis, and A.\ J.\ Tolley,
JHEP 08 (2021) 081 [arXiv:2012.09711 [hep-th]].


\bibitem{Li:2022rel}
Y.~Li, Y.-F.~Hang, and H.-J.~He,
JHEP 03 (2023) 254 [arXiv:2209.11191 [hep-th]].


\bibitem{Li:2021yfk}
Y.~Li, Y.-F.~Hang, H.-J.~He, and S.~He,
JHEP {02} (2022) 120 [arXiv:2111.12042 [hep-th]].


\bibitem{songhe}
T.\ Bargheer, S.\ He, and T.\ McLoughlin,
Phys.\ Rev.\ Lett.\ 108 (2012) 231601 [arXiv:1203.0562 [hep-th]].


\bibitem{ythuang}
Y.\ t.\ Huang and H.\ Johansson,
Phys.\ Rev.\ Lett.\ 110 (2013) 171601 [arXiv:1210.2255 [hep-th]].


\bibitem{DC-3dx}
N.\ Moynihan, JHEP 12 (2020) 163 [arXiv:2006.15957 [hep-th]].


\bibitem{DC-3dx2}
D.\ J.\ Burger, W.\ T.\ Emond, and N.\ Moynihan,
JHEP 01 (2022) 017 [arXiv:2103.10416 [hep-th]].


\bibitem{Gonzalez:2021bes}
M.~C.~Gonz\'alez, A.~Momeni, and J.~Rumbutis,
JHEP 08 (2021) 116 [arXiv:2107.00611 [hep-th]].


\bibitem{Moynihan:2021rwh}
N.~Moynihan,
arXiv:2110.02209 [hep-th].
	

\bibitem{Jackiw:1991}
R.\ Jackiw and V.\ P.\ Nair,
Phys.\ Rev.\ D\,43 (1991) 1933-1942.


\bibitem{Pisarski:1985}
R.~D.~Pisarski and S.~Rao,
Phys.\ Rev.\ D\,32 (1985) 2081.


\bibitem{UIR3}
B.~Binegar,
J.\ Math.\ Phys.\ {23} (1982) 1511.


\bibitem{supp}
Y.~F.~Hang and H.~J.~He,
Supplementary Material.

\bibitem{ET94}
H.\ J.\ He, Y.\,P.\,Kuang and X.\,Li,
Phys.\ Rev.\ D 49 (1994) 4842;
and Phys.\ Rev.\ Lett.\ 69 (1992) 2619.




\bibitem{ET96}
H.\ J.\ He and W.\,B.\ Kilgore,
Phys.\ Rev.\ D 55 (1997) 1515 [hep-ph/9609326].


\bibitem{weinbergPC}
Steven Weinberg, Physica 96A (1979) 327.
	
\bibitem{ETPC-97}
H.\ J.\ He, Y.\,P.~Kuang and C.\,P.~Yuan,
Phys.\ Rev.\ D\,55 (1997) 3038 [hep-ph/9611316];
 Phys.\ Rev.\ D 51 (1995) 6463 [arXiv:hep-ph/9410400].


\bibitem{Witten-3dTMG}
E.\ Witten,
``Three-Dimensional Gravity Revisited'',
arXiv:0706.3359 [hep-th];	
E.\ Witten,
``(2+1)-Dimen-sional Gravity as an Exactly Soluble System'',
Nucl.\ Phys.\ B 311 (1988) 46.

	
\bibitem{Hinterbichler-Rev}
For reviews, 
S.\ Carlip, ``Quantum Gravity in 2+1 Dimensions'',
Cambridge University Press, 1998; and
J.\ Korean Phys.\ Soc.\ 28 (1995) S447 [arXiv:gr-qc/9503024];
%
K.\ Hinterbichler,
Rev.\ Mod.\ Phys. 84 (2012) 671 [arXiv:1105.3735 [hep-th]].


\bibitem{Soldate:1986mk}
M.~Soldate,
Phys.\ Lett.\ B\,186 (1987) 321.


\bibitem{He2005}
D.\,A.\ Dicus and H.-J. He,
Phys.\ Rev.\ D\,71 (2005) 093009 [arXiv:hep-ph/0409131];
and Phys.\ Rev.\ Lett.\ 94 (2005) 221802 [hep-ph/0502178].	


\bibitem{sekhar}
R.\ S.\ Chivukula, D.\ Foren, K.\ A.\ Mohan, D.\ Sengupta, and E.\ H.\ Simmons,
Phys.\ Rev.\ D 101 (2020) 055013 
[arXiv:1906.11098 [hep-ph]];
Phys.\ Rev.\ D 101 (2020) 075013
[arXiv:2002.12458 [hep-ph]].


\bibitem{kurt}	
J.\ Bonifacio and K.\ Hinterbichler,
JHEP 1912 (2019) 165 [arXiv:1910.04767 [hep-th]].




\bibitem{Hang:2021oso}
Y.-F.~Hang, H.-J.~He, and C.~Shen,
JHEP {01} (2022) 153 [arXiv:2110.05399 [hep-th]].


\bibitem{S-Banerjee:2000gc}
R.~Banerjee, B.~Chakraborty and T.~Scaria,
Int.\ J.\ Mod.\ Phys.\ A\,16 (2001) 3967
[arXiv:hep-th/0011011 [hep-th]].

\end{thebibliography}
\end{document}